\documentclass[prd,aps,floats,twocolumn,nofootinbib]{revtex4-1}
\usepackage{slashed}
\usepackage{mathtools}
\usepackage{amsfonts}
\usepackage{amssymb}
\usepackage{epsfig}
\usepackage{rotating}
\usepackage{url}
\usepackage{times}
\usepackage{color}
\usepackage{bm}
\usepackage{xcolor,colortbl}
\usepackage{hyperref}
\usepackage[alsoload=hep]{siunitx}
\usepackage{enumitem}
\usepackage{fancyhdr}
\usepackage{anyfontsize}
\usepackage{wasysym}
\usepackage{empheq}

\newcommand{\At}{\ensuremath{\tilde{A}}}

\newcommand{\grad}{\ensuremath{\vec{\nabla}}}
\newcommand{\Order}{{\cal O}}

\newcommand{\ph}{\phantom}
\newcommand{\nn}{\nonumber}

\newcommand{\Mb}{{\bf M}}

\newcommand{\sigmaQui}{ \sigma^{(5)}}
\newcommand{\Tr}{{\rm Tr}}

\newcommand{\tn}{\tilde{\nabla}}
\newcommand{\squareE}{\ensuremath{\tilde{\square}}}
\newcommand{\squareM}{\ensuremath{\square}}
\newcommand{\connE}{\ensuremath{\tilde{\nabla}}}
\newcommand{\connM}{\ensuremath{\nabla}}

\newcommand{\metM}{\ensuremath{g}}

\newcommand{\volM}{\ensuremath{\sqrt{-\metM}}}
\newcommand{\RiemE}{\ensuremath{\tilde{R}}}
\newcommand{\RiemM}{\ensuremath{R}}

\newcommand{\vp}{\varphi}
\newcommand{\tx}{\tilde{X}}

\newcommand{\ven}{\vec{\nabla}}

\definecolor{orange}{rgb}{1,0.5,0}
\definecolor{darkorange}{rgb}{0.69,0.33,0.13}
\definecolor{fidcol}{rgb}{0.7,0,0}

\begin{document}

\title{The Parameterized Post-Newtonian-Vainshteinian formalism for the Galileon field}

\author{Nadia~Bolis$^{1}$}
\email{bolis@fzu.cz, nbolis@ucdavis.edu}
\author{Constantinos~Skordis$^{1,3}$}
\email{skordis@ucy.ac.cy}
\author{Daniel B~Thomas$^{2}$}
\email{daniel.thomas-2@manchester.ac.uk}
\author{Tom~Z\l o\'{s}nik$^{1}$}
\email{zlosnik@fzu.cz}

\affiliation{
	$^1$ CEICO, Institute of Physics of the Czech Academy of Sciences, Na Slovance 1999/2, 182 21, Prague \\
	$^2$ Jodrell Bank Centre for Astrophysics, School of Physics and Astronomy,
	The University of Manchester, Manchester M13 9PL, UK\\
	$^3$ Department of Physics, University of Cyprus, 1, Panepistimiou Street, 2109, Aglantzia, Cyprus 
}

\begin{abstract}
Recently, an extension to the Parameterized Post-Newtonian (PPN) formalism has been proposed. This formalism, the Parameterized Post-Newtonian-Vainshteinian (PPNV) formalism, is well suited to theories which exhibit Vainshtein screening of scalar fields. In this paper we apply the PPNV formalism to the Quartic and Quintic Galileon theories for the first time. As simple generalizations of standard scalar-tensor field theories they are important guides for the generalization of parameterized approaches to the effects of gravity beyond General Relativity. In the Quartic case, we find new PPNV potentials for both screened and un-screened regions of spacetime, showing that in principle these theories can be tested.
In the Quintic case we show that Vainshtein screening does not occur to Newtonian order, meaning that the theory behaves as Brans-Dicke to this order, and we discuss possible higher order effects.
\end{abstract}

\maketitle

%%%%%%%%%%%%%%% Outline %%%%%%%%%%%%%%%%%%%%%%%%%%%%%%
\section{Introduction}
Due to its dramatic success in explaining observations ranging from table-top tests of the gravitational force to strong-gravity environments such as systems of merging black holes, General Relativity remains the preferred theory of gravity. Its success on astrophysical and cosmological scales is perhaps less clear however; here there is considerable evidence for a dark sector in the universe, comprised of dark matter and dark energy. Though the dark matter may represent new particle physics and the dark energy may be a cosmological constant, it is conceivable that the evidence for either or both arises from a modification to gravity.

The question that arises then is how to discriminate between modified theories of gravity and General Relativity. There are a great number of alternative gravitational theories which could be tested, so many in fact that it would be inefficient to test each one separately. A more effective approach is to construct parameterized frameworks which can be applied to certain gravitational systems. From the theoretical side, one can deduce 
values of the framework's parameters that given theories predict; then, from the experimental side, data from these gravitational systems can be used to put constraints on these parameters. In this manner, whole sets of theories can be constrained and even excluded.

Early examples of such parameterized frameworks became eventually known as the Parameterized Post-Newtonian (henceforth PPN) 
framework \cite{Will1971,WillNordtvedt1972,Will2014}.  As the name suggests, this formalism parameterizes the gravitational field 
beyond the limit of Newtonian gravity. Specifically, it is assumed that the gravitational field is described - at least in part - 
by a metric tensor and that throughout a system such as the solar system, the metric tensor can be described as a slightly 
perturbed Minkowski spacetime.  These perturbations are parameterized in terms of a set of potentials (the PPN potentials), 
each of which are expressible as spatial integrals over components of the matter stress energy tensor. Crucially, it is then 
possible to relate constant coefficients multiplying these potentials to important observables in the solar system. Data from 
lunar laser ranging and the motion of bodies within the solar system has significantly constrained many of these constants and in doing so 
has severely constrained a number of alternatives to General Relativity \cite{Will2014}. 
Other parameterized frameworks have been developed over the years~\cite{Damour:1991rd, Freire2010,Yunes:2009ke, Cornish:2011ys, Loutrel:2014vja, Sampson2013,Zhang:2007nk,Amin:2007wi, Silvestri:2013ne,Hu:2007pj, Hu:2008zd,Skordis:2008vt, Baker:2011jy, BakerFerreiraSkordis2013,Battye:2013aaa, Battye:2012eu,Bloomfield:2012ff, Gubitosi:2012hu, Piazza:2013coa}, adapted to various systems of interest from the strong-field to cosmology.

If we take the stress energy of matter to be that due to visible matter in the solar system, it's clear that the PPN formalism will be limited 
if the theory of gravity is such that the spacetime metric cannot be expressed in terms of the regular PPN potentials 
(see for example those listed in Box 2 of \cite{Will2014}).  This can indeed occur, and happens typically in modified theories of gravity that 
introduce additional scales into gravitational physics.  A simple example is that of a scalar field of mass $m$ coupled to matter 
in a manner that gives~\cite{Wagoner:1970vr} a Yukawa-type `$e^{-mr}/r$' contribution to the gravitational field that test particles feel; 
this contribution is not covered by the regular PPN potentials, although for small 
enough $m$ a perturbative approach in terms of these potentials can give sufficient accuracy~\cite{Wagoner:1970vr,WillBook}.

A particularly interesting family of scalar-tensor gravitational theories are the Galileon theories \cite{NicolisRattazziTrincherini2008}, which introduce 
a scalar field $\chi$ into the gravitational sector. These theories are a special subset of Horndeski scalar-tensor theories that 
possess field equations with no higher than second-order time derivatives and an emergent Galilean symmetry of the Lagrangian - up to total derivatives - 
under the transformation  $\partial_{\mu}\chi\rightarrow \partial_{\mu}\chi+ v_{\mu}$ in Minkowski spacetime. These theories have attracted much 
attention as a potential candidate for dynamical dark energy \cite{Clifton:2011jh}.  The most general Galileon Lagrangian consists of five 
independent terms: the first two are a term simply proportional to the Galileon field $\chi$ and a canonical kinetic term for $\chi$. The 
remaining three terms are non-canonical kinetic terms for $\chi$, respectively referred to as the Cubic, Quartic, and Quintic Galileon (named 
after the order at which $\chi$ appears in their Lagrangians e.g. the Cubic Galileon is cubic in $\chi$).

For dimensional reasons, the non-canonical Galileon kinetic terms involve dimensionful constants and hence introduce a new scale into gravitation, an
energy scale $\Lambda$.
 For example, consider the case where the Galileon sector is described by a canonical kinetic term alongside a Cubic Galileon piece. It is known that in static  spherical symmetry situations (and assuming conformal coupling to matter) there is asymptotically a Brans-Dicke-type fifth force due to the canonical kinetic term at very large distances from the central gravitating matter source. Remarkably though, as one moves to smaller radii, the non-linear contribution of the Cubic Galileon term becomes more and more important. At distances from the source much smaller than a certain radius $r_{V}$ (a scale built from the mass $M$ of the source, the Planck mass, and the dimensionful coefficient of the Cubic Galileon term), the profile of the scalar field is dominated by these non-linear terms and leads to a dramatic suppression of the fifth-force relative to the Brans-Dicke form. The scale $r_{V}$ is referred to as the \emph{Vainshtein radius} and the suppression of the fifth-force is referred to as \emph{Vainshtein screening}. The existence of Vainshtein screening is important for the phenomenological viability of these models: it provides a simple way for the theory to have a relatively dominant effect on late-time, large-scale cosmology whilst having a sufficiently small effect on gravity in the solar-system to have avoided exclusion by experiment.

Since the detection of the neutron star-neutron star merger event GW170817 \cite{TheLIGOScientific:2017qsa, Monitor:2017mdv} and the resulting constraint on the gravitational wave speed the Quartic and Quintic Galileon theories are no longer strong candidates for explaining Dark Energy \cite{Baker2017,Ezquiaga2017,SaksteinJain2017,CopelandEtAl2018}.
However, the scale $\Lambda$ may be made large enough so that such theories no longer play the role of Dark Energy 
and the gravitational wave constraints need not apply. As such, the results presented here serve as an important guide in constraining theories which deviate from General Relativity (GR) in the Infra-Red (IR), using strong field data.

As in the case of Yukawa-type modifications to a scalar field profile, the effect of Vainshtein screening is not covered by the PPN potentials. It is necessary then to modify the PPN formalism to introduce a parameterization of fields that is sufficiently general to account for the presence of Vainshtein screening. Such a proposal was put forth in \cite{AvilezEtAl2015} and is termed the Parmeterized Post-Newtonian Vainshteinian formlism (henceforth PPNV).
The authors applied their formalism to the case where the Galileon sector consisted of a scalar field with a canonical kinetic term alongside a Cubic Galileon term. There are important benefits from the development of such a formalism:
\begin{enumerate}
\item Gravitational physics in the solar system, by and large, lacks high symmetry in space and time and the field equations of General Relativity are non-linear. The PPN formalism, as a perturbative formalism, helps systematically break the full equations into easier-to-solve sets of equations. This is similarly true for the gravitational (including scalar field) equations in the PPNV formalism.
\item As in the PPN formalism, the parameterized nature of the PPNV formalism may point towards design of experiments to most accurately probe the effects of a field such as the Galileon i.e. they should be experiments that most directly constraint PPNV coefficients.
\item The Vainshtein screening mechanism and behavior in the non-linear regime is currently largely understood in examples of high symmetry. The apparent accuracy of the perturbative approach in a given, less symmetrical situation may yield insight into the distribution of screened and non-screened regions through spacetime.
\end{enumerate}
The layout of the paper is as follows: In Section \ref{PPNandPPNV} we provide a brief technical overview to the structure of PPN and PPNV formalisms. In Section \ref{cubic} we discuss the earlier application of the PPNV formalism to the case of the Cubic Galileon theory. In Sections \ref{quartic} and \ref{quintic} we proceed to apply the PPNV formalism to both Quartic and Quintic Galileon theories. Finally in Section \ref{discussion} we discuss our results and present our conclusions. Throughout the article we use units such that the speed of light is unity.

\section{PPN and PPNV Review}
\label{PPNandPPNV}
We now present a brief overview of the PPN and PPNV formalisms. In the PPN formalism it is assumed that one of the constituents of the gravitational field is a metric tensor $g_{\mu\nu}$ and that this tensor is approximately the metric tensor $\eta_{\mu\nu}$ of Minkowski space plus a small correction $h_{\mu\nu}$:

\begin{align}
g_{\mu\nu} = \eta_{\mu\nu} + h_{\mu\nu}
\label{metric_eta_h}
\end{align}
Throughout we will use units where $c=1$. 
This metric ansatz is valid as long as the time dependence of the background metric and scalar field are sufficiently small and as long as we are concerned with situations away from compact objects, such as black holes, or concerned  with systems which have reached a type of quasistatic equilibrium.
Though this is clearly not an accurate description of our entire universe, this ansatz describes the geometry of the solar system to a good approximation \cite{IshibashiWald2005}.
\footnote{Indeed, typically constraints have arisen on the Galileon theories from cosmological data where time variation in the `background' metric is important \cite{BarreiraEtAl2014} }. 
The matter content is assumed to take the form of a fluid (potentially with anisotropic stresses) and it is assumed that the stress energy tensor of matter is covariantly conserved with respect to the covariant derivative $\nabla_{\mu}$ associated with $g_{\mu\nu}$. The velocity of matter $v^i$ is observed to be typically of order $\sim 10^{-3}-10^{-4}$ in units where the speed of light is unity, and this is taken to be the leading order of smallness in the PPN expansion (i.e. the $v^{i}$ are allocated PPN order ${\cal O}_{PPN}(1)$). By the allocation $v^{i}\sim {\cal O}_{PPN}(1)$, then $v^{i}\sim |d/dt|/|d/dx|$ and so time derivatives are taken to increase the PPN order whilst spatial derivatives do not. 
Typical Newtonian potentials are of order $\sim v^{2}$ and so are allocated PPN order ${\cal O}_{PPN}(2)$, whilst typical matter densities - via the assumed approximate validity of Poisson's equation in the Newtonian limit - are also of ${\cal O}_{PPN}(2)$. 

The full equations of the system are taken to be the gravitational field equations (describing the dynamics of $g_{\mu\nu}$ as well as other gravitational fields that may exist such as a scalar field $\phi\equiv e^{-2\chi/M_{P}}$ 
in the case of scalar-tensor theory) and matter field equations (that may be recovered from equations of energy-momentum conservation). Using the above order allocations, one can proceed to perturbatively expand the full equations to order ${\cal O}_{PPN}(2)$ (Newtonian limit) and ${\cal O}_{PPN}(>2)$ (post-Newtonian corrections). The PPN formalism has been applied to a wide variety of theories such as Brans-Dicke theory \cite{Will2014} and the Einstein-Aether theory \cite{FosterJacobson2005}.

Now we turn to the case of the Galileon theory. 
In the simplest example of a Galileon with non-canonical kinetic terms, it is known in the Vainshtein screening region that there is a correction $\delta_{V}U$ to the Newtonian gravitational potential due to a spherically symmetric source which goes approximately as $\delta_{V}U\sim U\times(r/r_{V})^{3/2}$ \cite{ChowKhoury2009}, where $U$ is the canonical Newtonian potential. It is clear by inspection of the form of the PPN potentials \cite{Will2014} that this correction cannot be constructed from linear combinations of these potentials. It is necessary then to \emph{extend} the PPN formalism to include potentials of which the above correction is an example.

The formalism proposed in \cite{AvilezEtAl2015} is an example of such an extension. The idea is to add an \emph{additional} order in the expansion of fields that quantifies the effect on fields due to proximity to the boundary between regions with and without Vainshtein screening. From the above example of the Cubic Galileon, one can imagine that in solving the full equations that the contribution to the gravitational potential may go as $ \delta_{V} U\sim U\times\left((r/r_{V})^{3/2} +  {\cal O}((r/r_{V})^{3})+\dots\right)$.  In this example then it seems reasonable to assign a \emph{Vainshteinian order} $V$ to terms which have a dependence $(r/r_{V})^{3V/2}$ and retain the PPN order $N$ for remaining dependencies.  How do we assinged orders to the various quantities for a general theory?

Following \cite{AvilezEtAl2015}, let's consider a general setting where the theory in question containts an additional scalar field with
 non-canonical kinetic terms in the action, leading to Vainshtein-type effects. As in \cite{AvilezEtAl2015} we are also concerned with theories leading 
to a single Vainshtein scale. Typically, and assuming the scalar $\chi$ has dimensions of mass, 
outside the Vainshtein radius the scalar field solution will be dictated by the canonical term to be $\chi \sim (M/M_P) r^{-1}$ where $M$ is the mass of the source.
Deep inside the Vainshtein radius, the non-linear interactions will switch on so that the scalar equation schematically  reads as
\begin{equation}
  \frac{\partial^{m} \chi^{n-1}}{\Lambda^{m + n - 4}} \sim \frac{\rho}{M_P}
\end{equation}
where $\rho$ is the matter density and $\Lambda$ is the strong-coupling scale. Here $m$ is an integer specifying the number of derivatives and
 $n$ another integer specifying the how many occurrences of $\chi$ appear in the action for the term in question. Thus this prescription leads to
a Vainshtein scale where classical perturbation theory breaks down given by $r_V \Lambda \sim  (M/M_P)^s$ where $s = \frac{n-2}{m+n-4}$. In spherically symmetric
situations, to lowest Newtonian order deep inside the Vainshtein radius, $\chi$ will schematically take the form
\begin{equation}
 \frac{\chi}{M_P} \sim \left(\frac{r_S}{r}\right) \left(\frac{r}{r_V}\right)^{k V}
\end{equation}
where $r_S$ is the Schwarzschild radius of the source and
\begin{equation}
 k = \frac{m+n-4}{n-1} 
\end{equation}
is a fraction fixed by the action of the theory under consideration. In the case of the cubic galileon, $m=4$ and $n=3$ so that $k=3/2$.

As such - and using the notation ${\cal O}_{PPNV}(N,V)$ to denote a quantity of PPN order $N$ and Vainshteinian order $V$  - we allocate the following \emph{PPNV} orders to the contributions to the Newtonian potential:
\begin{align}
U &\sim {\cal O}_{PPNV}(2,0) \nn\\
U\times(r/r_{V})^{k}  &\sim {\cal O}_{PPNV}(2,1) \nn\\
U\times (r/r_{V})^{2k }  &\sim {\cal O}_{PPNV}(2,2) \nn\\
\end{align}

In general, we assign an PPN order $N$ and Vainshteinian order $V$ to any operator in the theory under consideration using the prescription
\begin{equation}
 {\cal O}_{PPNV}(N,V) \sim  r_s^{N/2} r_V^{-kV} 
\end{equation}

\begin{figure}
	\center
	\epsfig{file=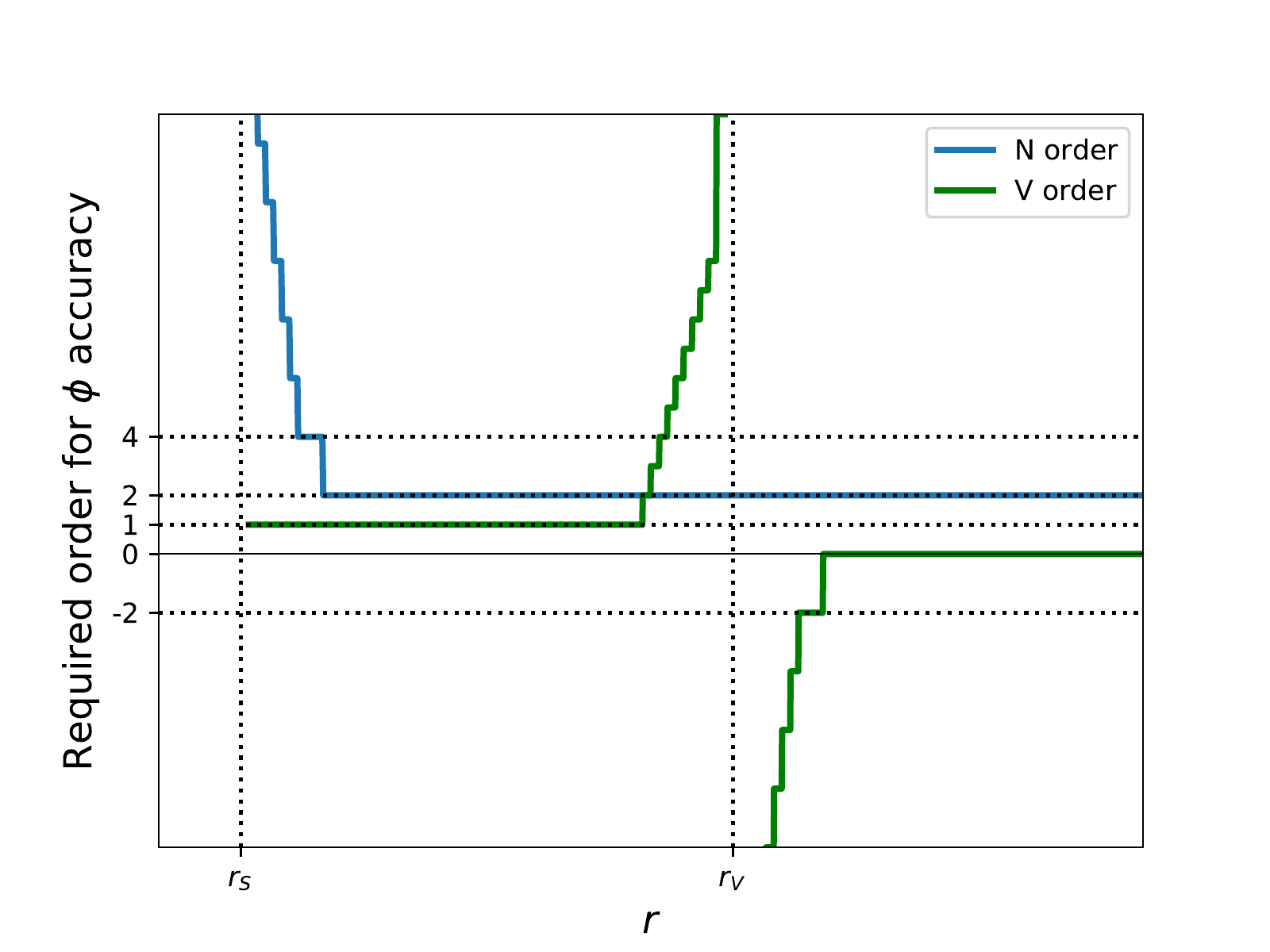,width=9.cm}
	\caption{Schematic illustration for the Cubic Galileon case of the PPNV order in the scalar field $\phi$ that one should go to, to achieve a certain standard of approximation to the results from solving the full equations exterior to a black hole with Schwarzschild radius $r_{S}$ and Vainshtein radius $r_{V}$ (see \cite{BabichevEtAl2016} for relevant cases). The non-continuity of the $V$ Order curve either side of $r_{V}$ illustrates the need for use of a dual formulation of the theory on the inside regions}
	\label{orders}
\end{figure}
Figure \ref{orders} schematically illustrates in the case of the Cubic Galileon, how $N$ and $V$ orders of greater and greater magnitude are expected to be necessary in describing, to a set accuracy, the scalar field profile exterior to a black-hole type solution with Schwarzschild radius $r_{S}$; it can be seen that, as expected, greater and greater PPN orders $N$ are required to account for the fact that more and more orders are needed to account for increasingly post-Newtonian behavior as one approaches the event horizon. Close to the Vainshtein screening barrier it can be seen that more and more $V$ orders are necessary as $r/r_{V} \rightarrow 1$; the reason that the green steps are not continuous on either side of the boundary is because leading corrections outside the screened region go as positive powers of $(r_{V}/r)^{3/2}$. This is not to say that somehow the `weak field limit' may no longer be applicable here but that the presence of Vainshtein screening means that power-law expansions in the orders $(N,V)$ cannot cover both screened and un-screened regions together.

The question of how to allocate a $V$ order to a quantity seems quite clear in spherical symmetry - as proximity to the Vainshtein screening boundary is measured by $r/r_{V}$, but how does one do this when the geometry of the Vainshtein screening boundary may be more complicated due to less-symmetric mass distributions? By comparison to the PPN formalism, it can be noted that in spherical symmetry the Newtonian potential goes as $r_{S}/r$ - where $r_{S}$ is the Schwarzschild radius of the source and the post-Newtonian correction to this potential goes as $(r_{S}/r)^{2}$; though the PPN expansion is not an expansion in $r_{S}$, there is a correspondence here between powers of $r_{S}$ appearing in potentials and PPN order $N$. The approach of the PPNV formalism is to pre-empt an extension of this in the Vainshteinian case by \emph{assigning} a PPNV order to the dimensionful constant appearing in the non-canonical kinetic term of the Galileon action. We will now attempt to make this approach clearer by seeing its application in the case of the Cubic Galileon theory.

\section{Cubic Galileon review}
\label{cubic}
In this section we briefly discuss the Cubic Galileon and the application of the PPNV formalism to it. This case is discussed in more detail in \cite{AvilezEtAl2015} and we review the authors' results here. The Cubic Galileon theory has one additional parameter beyond the standard scalar-tensor theory, namely, the scale $\Lambda$.
 It has the following action:

\begin{align}
S_{3}[\tilde{g},\chi] &= \frac{1}{16\pi G}\int d^{4}x \sqrt{-\tilde{g}}
\tilde{R} \nn\\
&+ \int d^{4}x \sqrt{-\tilde{g}} \bigg(c_{0}\tilde{X} + \frac{1}{\Lambda^{3}}\tilde{X}\tilde{\square}\chi\bigg)  + S_{M}[g] \label{cubic_action}
\end{align}
where for notational compactness the functional dependence of the action on matter fields is not written and 

\begin{align}
\tilde{X}  &\equiv  -\frac{1}{2}\tilde{g}^{\mu\nu}\tilde{\nabla}_{\mu}\chi\tilde{\nabla}_{\nu}\chi 
\label{tildexdef} 
\\
g_{\mu\nu} &=  e^{2\chi/M_{P}}\tilde{g}_{\mu\nu}
\end{align}
where $M_{P}\equiv 1/\sqrt{8\pi G}$. Note that the action above is shift symmetric in $\chi$.
As discussed in Section \ref{PPNandPPNV}, it is a convention in the PPN formalisms and some extensions thereof to work with the metric that is minimally coupled to matter, that is, $g_{\mu\nu}$. Therefore it is useful to write the action (\ref{cubic_action}) entirely in terms of $g_{\mu\nu}$ and a scalar field and it is useful to work with the scalar field  $\phi\equiv e^{-2\chi/M_{P}}$; then, (\ref{cubic_action}) can be written, up to a boundary term, as:

\begin{align}
S[g,\phi] &=  \frac{1}{16\pi G}\int d^{4}x\sqrt{-g}\bigg[\phi R + \frac{2\omega}{\phi}Y-\frac{\alpha_{3}}{4}\frac{Y}{\phi^{3}}\square\phi\bigg] \nn\\
 & + S_{M}[g] 
\label{cubicmatterframe_action}
\end{align}
where 

\begin{align}
\alpha_{3} &\equiv \frac{M_{P}}{\Lambda^{3}}\\
\omega &\equiv  \frac{c_{0}-6}{4} \label{omegadef} \\
Y & \equiv -\frac{1}{2}g^{\mu\nu}\nabla_{\mu}\phi\nabla_{\nu}\phi \label{ydef}
\end{align}
It can be shown that in the static, spherically symmetric weak field limit \cite{ChowKhoury2009}, in the exterior a mass $M$ there exists screening for $r\ll r_{V}$ where $\alpha_3$ is related to $r_{V}$ and the mass's Schwarzschild radius $r_{S}$ as:

\begin{align}
\alpha_{3} &= \frac{1}{4\pi}\frac{r_{V}^{3}}{r_{S}}
\end{align}
Given the discussion in Section \ref{PPNandPPNV}, this suggests that it can be important to assign a PPNV order to the dimensionful number $\alpha_{3}$. If (post-)Newtonian effects of order ${\cal O}_{PPN}(N)$ in spherical symmetry appear in the combination $r_{S}^{N/2}$ and Vainshteinian effects appear in the combination $r_{V}^{3V/2}$.
Under our convention, it is clear that $\alpha_3$ gets assigned the PPNV orders 

\begin{align}
{\cal O}_{PPNV}(\alpha_{3}) &=  (-2,-2)  \label{alphaorder}
\end{align}
This is an important part of constructing frameworks beyond PPN: whereas in PPN it was only necessary to assign values to fields such as the metric - or the matter density field - when there exist fixed, new scales in the problem (as for the parameter $\alpha_{3}$), one can meaningfully associate perturbative orders to these scales. 

Given the action (\ref{cubicmatterframe_action}) one can then 
obtain the equations of motion for matter and gravitational fields (here taken to be $g_{\mu\nu}$ and $\phi$) and these may be found in~\cite{AvilezEtAl2015}. 
As is suggested by Figure \ref{orders}, 
care must be taken when expanding fields in $V$ order depending on whether one is within a screened region or not - specifically that an 
expansion in `opposite' powers of $V$ is necessary in each region. It is argued in \cite{AvilezEtAl2015} that in an unscreened (henceforth \emph{outside}) 
region, the appropriate expansion for $\phi$ is:
\begin{align}
\phi^{(out)} = \phi_0^{(out)}\left[1+\sum_{N=2}^{\infty}\sum_{V=0}^{-\infty}\varphi^{(N,V)}(t,x^{i})\right].
 \label{phianzout}
\end{align}
where $\phi_0^{(out)}$ corresponds to the value of the scalar field as $r\rightarrow \infty$ and is taken to have PPNV order $(0,0)$.

Performing the PPNV expansion in the inside region necessitates the use of auxiliary fields which are dual to the interaction terms 
determined by $\nabla_\mu \phi$.  In the original PPNV article~\cite{AvilezEtAl2015}, this was achieved via a 
Legendre transformation of the action using the dualization procedure of~\cite{Padilla:2012ry}. Dualizing the action leads to the 
absorption of powers of the parameter $\alpha_3$ into the auxiliary fields, 
such that the action becomes perturbative in inverse powers of $\alpha_3$.  The field equations for the auxiliary fields 
(denoted by $A_\mu$ and $Z$ in ~\cite{AvilezEtAl2015}, dual to $\nabla_\mu \phi$ and $\square \phi$ respectively) 
were obtained from the dual action, and those field equations were subsequently expanded using the PPNV formalism.

The above procedure is fairly lengthy but fortunately it is not necessary; one can obtain the desired field equations in the inside region  
directly from the ones relevant to the outside region without first dualising the action. The trick is to specify an auxiliary field $B$, 
related to $\phi$ by an appropriate rescaling via a power of $\alpha_3$, a method essentially identical to the one 
used in \cite{McManus:2016kxu,McManus:2017itv}, although that formalism also encompasses chameleon fields.
To be more precise, we define 
\begin{equation}
%\nabla_\mu \phi = \phi_0^{in)}   \alpha_3^n\nabla_\mu B
\nabla_\mu \phi =    \alpha_3^p\nabla_\mu B
\label{phi_to_B_cubic}
\end{equation}
 for an unknown power $p$.
%, where we have chosen to normalize $B$ by the $\Order(0,0)$ constant $\phi_0^{(in)}$ (the reason for this normalization is clarified further below). 
 This is then inserted into the Galileon field equations,
 replacing all occurances of $\nabla_\mu \phi$, but leaving $\phi$-dependent terms without derivatives intact.
Following that, one chooses the power $p$ such that there is at least one term without any $\alpha_3$ (this is the leading order term) 
while all other terms have powers of $\alpha_3$ with strictly negative exponent. 
In this way, one can safely take the limit $\alpha_3\rightarrow \infty$ which is the deep Vainshtein limit. 

In the case of the Cubic Galileon, $p=-1/2$, so that $\nabla_\mu \phi = \alpha_3^{-1/2} \nabla_\mu B$ and the covariant dual field equations 
which contain  both $\phi$ and $B$  can be found in \cite{AvilezEtAl2015}. From \eqref{phi_to_B_cubic} we may then relate the two fields so that
\begin{equation}
%\phi^{(in)} = \phi_0^{(in)} \left( 1 + \alpha_3^{-1/2} B \right)
\phi^{(in)} =  1 + \alpha_3^{-1/2} B 
\label{phianzin}
\end{equation}
which makes the choice of the normalization in \eqref{phi_to_B_cubic} clear. 
We have specifically normalized $\phi^{(in)}$ such that it tends to unity when all Vainshteinian corrections subside
  (e.g. as $r\rightarrow 0$ in the case of spherical symmetry) and this has the outcome that the bare gravitational parameter $G$ 
can be identified with Newton's constant.
%  Moreover, the constant $\phi_0^{(in)}$ clearly corresponds to the value of $\phi$ as $r\rightarrow 0$.  
Notice that the constant $\alpha_3^{-1/2}$ is of order $\Order_{PPNV}(1,1)$ meaning that it increases both the 
PPN order and the Vainshtein order of any terms multiplying it by one. 

The field $B$ also has $N$ and $V$ orders; in the case of Cubic Galileon
\begin{align}
B  =   \sum_{N=1}^{\infty}\sum_{V=0}^{\infty} B^{(N,V)}(t,x^{i})
\label{Banz}
\end{align}
so that the leading order is  $B^{(1,0)}$.

The reason that one looks at an expansion in positive powers of $V$ in an inside region and negative powers of $V$ in an outside region is a reflection of our convention for what $V$ is: by choice, the quantity $\alpha_{3}$ has been allocated a negative $V$ order and it is anticipated that $\alpha_{3}$ will
 appear in negative powers in a perturbative expansion of $\phi^{(in)}$; thus terms will have a non-negative $V$ order. 
The opposite is true in an outside region, where $\alpha_{3}$ will appear in positive powers in a perturbative expansion of $\phi^{(out)}$, 
and hence these terms will have a negative $V$ order.

It is expected that both of these perturbative expansions will cease to work close to any boundary between inside and outside regions.
Indeed, in the absence of knowing the distribution of such boundaries in space - given a certain matter content in space - the breakdown of such an expansion may be a sign of approach to a boundary (for example if at some point within an outside region, $\phi^{(2,-2)}$ becomes as significant as $\phi^{(2,0)}$ etc.).

We now briefly summarize the results obtained for inside and outside regions in the Cubic Galileon case.

\subsection{Outside Region}
It is found that up to $N$ order $2$ and $V$ order $-2$ that the scalar field perturbations $\varphi$ and non-vanishing parts of $h_{\mu\nu}$ are:
\begin{align}
\varphi^{(2,\leq|-2|)} &= \frac{2G}{(3+2\omega)\phi_0^{(out)}}U
\nn
\\
&-\frac{\alpha_{3}}{8(3+2\omega)\phi_0^{(out)}}\left(\frac{G_{C}}{2+\omega}\right)^{2}U^{(out,3)}_{V_{1}}
 \label{cubicoutphi}
\\
h_{00}^{(2,\leq|-2|)} &= 2G_{C}U  +2g^{(out,3)}_{V_{1}}\left(\frac{M_{P}}{H_{0}}\right)^{2}G_{C}^{3}U^{(out,3)}_{V_{1}} 
\label{cubicouthoo} 
\\
h_{ij}^{(2,\leq|-2|)} &= \left[2\gamma G_{C}U +2\gamma^{(out,3)}_{V_{1}}\left(\frac{M_{P}}{H_{0}}\right)^{2}G_{C}^{3}U^{(out,3)}_{V_{1}}\right]\gamma_{ij} 
\label{cubicouthij}
\end{align}
where $\gamma_{ij}$ is the metric of flat three-dimensional Euclidean space in arbitrary coordinates and where first and second terms for each field/field component  are of $(N,V)$ order $(2,0)$ and $(2,-2)$ respectively and where 
the potentials $U$ and $U^{(out,3)}_{V_{1}}$ are
\begin{align}
U(x) &\equiv \int \frac{\rho(\vec{x}',t)}{|\vec{x}-\vec{x}'|}d^{3}x'
 \label{usol}
\\
U^{(out,3)}_{V_{1}}(x) &\equiv\int d^{3}x' d^{3}x''\rho(t,\vec{x}')\rho(t,\vec{x}'')\bigg[
\frac{(\vec{x}-\vec{x}')\cdot(\vec{x}-\vec{x}'')}{
	|\vec{x}-\vec{x}'|^{3}|\vec{x}-\vec{x}''|^{3}} 
\nn 
\\
&
  -2\frac{(\vec{x}-\vec{x}')\cdot(\vec{x}'-\vec{x}'')}{|\vec{x}-\vec{x}'|^{3}|\vec{x}'-\vec{x}''|^{3}} \bigg] 
\end{align}
The constant  $G_C$ 
is the `cosmological value' of the gravitational strength, i.e. the one that could be measured by cosmological probes. This value is not necessarily the same as the locally measured value (which for these theories will typically be determined by the specific form of the metric in the \emph{inside} region). For this particular theory $G_{C}$ is given by
\begin{equation}
G_{C} \equiv  \frac{(4+2\omega)}{(3+2\omega)}\frac{G}{\phi_0^{(out)}} 
\label{G_C}
\end{equation}
which is identical to the case of Brans-Dicke theory. This was expected as this Cubic Galileon theory asymptotes to Brans-Dicke far away from massive sources, as is relevant to cosmological scales.

The constant parameter $\gamma$ is one of the standard PPN parameters and for this particular theory is given by
\begin{equation}
\gamma \equiv  \frac{1+\omega}{2+\omega} 
\label{gamma_PPN}
\end{equation}
which is  identical, once again, to the case of Brans-Dicke theory.

%It can be seen then that in the outside region, the leading (${\cal O}_{PPNV}(2,0)$) effect of the scalar field is as in Brans-Dicke theory 
%where the scalar field is proportional to the Newtonian potential $U$ and the effect of this field on the tensor field $h_{\mu\nu}$ 
%is to rescale $G$ into $G_{C}$ and via factors of $\omega$ which also changes the relative sizes of $h_{00}$ and $\frac{1}{3}h^{i}_{\ph{i}i}$. 
%To next order (${\cal O}_{PPNV}(2,-2)$), the scalar field contributes effects via the new potential $U^{(out,3)}_{V_{1}}$. 

The  constant parameters $g^{(out,3)}_{V_{1}}$ and $\gamma^{(out,3)}_{V_{1}}$ are new parameters beyond the standard PPN. They are PPNV parameters
which measure the strength of the contribution of the Vainshteinian potential $U^{(out,3)}_{V_{1}}$ to the metric. 
While in General Relativity and in Brans-Dicke theory they are exactly zero, in this particular Cubic Galileon theory we have that
\begin{align}
 \gamma^{(out,3)}_{V_{1}} &\equiv \frac{\pi}{4}\left[\frac{M_{P}}{(2+\omega)\Lambda}\right]^{3}\left(\frac{H_{0}}{M_{P}}\right)^{2}
 =  \frac{\pi H_{0}^2}{4(2+\omega)^3}  \, \alpha_3
\end{align}
and $g^{(out,3)}_{V_{1}} = - \gamma^{(out,3)}_{V_{1}}$.
It is immediately observed that as $\alpha_3\rightarrow 0$ the theory asymptotes to Brans-Dicke, which could have been expected by inspection 
of the action \eqref{cubicmatterframe_action}.

The constant $H_{0}$ is taken to be the value of the Hubble parameter today. 
Note that the parameter $ \gamma^{(out,3)}_{V_{1}}$ is not the same as the parameter $\gamma_{V}$ from \cite{AvilezEtAl2015}. 
The number $H_{0}$ is incorporated into the definitions of $g^{(out,3)}_{V_{1}}$ and $\gamma^{(out,3)}_{V_{1}}$ so that for values of $\Lambda$ 
associated with the Galileon playing the role of dark energy (where it takes values ${\cal O}((M_{P}H_{0}^{2})^{1/3})$ \cite{AvilezEtAl2015}), 
that the value of $g^{(out,3)}_{V_{1}}$ and $\gamma^{(out,3)}_{V_{1}}$ is of order $(2+\omega)^{-3}$. 
What is a typical size for the parameter $\omega$? If it is the case that cosmologically the theory (\ref{cubicmatterframe_action}) is essentially 
that of Brans-Dicke theory \cite{Brans1961} (i.e. the influence of non-linear terms in the action have negligible effect) 
then we may use cosmological constraints on that theory restrict $\omega \gtrsim 1000$ \cite{AvilezSkordis2014} 
and so $\{g^{(out,3)}_{V_{1}} , \gamma^{(out,3)}_{V_{1}} \} \ll 1$.

The potentials $U$ and $U^{(out,3)}_{V_{1}}$ involve spatial integrals. The domain of integration for formal solutions to fields in the outside regime 
is understood to encompass all space, including inside and outside regions. Nonetheless, there remains a technical challenge in determining the distribution of inside and outside regions: one could imagine that the shape of these regions in general situations may be rather complex. How to find where the boundaries are? A possible strategy is to begin at a point in space where one can be reasonably confident one is in an outside regime (for example very far from matter sources). One can then compute physical effects of the scalar field to ${\cal O}_{PPNV}(2,-2)$ and ${\cal O}_{PPNV}(2,-4)$. If the ${\cal O}_{PPNV}(2,-4)$ effects are subdominant then one can have some confidence that one is in the outside region and that the PPNV expansion is appropriate. Then one can try to move to nearby points in space, mapping out boundaries demarcated by extensions where ${\cal O}_{PPNV}(2,-4)$ terms would give comparable effects to ${\cal O}_{PPNV}(2,-2)$ - this may indicate proximity to regions where the perturbative expansion breaks down (and thus, a transition to a screened region).

Finally, a word on notation is in order. The ``$(out,3)$'' in $U^{(out,3)}_{V_{1}}$ and also in $g^{(out,3)}_{V_{1}}$ and $\gamma^{(out,3)}_{V_{1}}$ denotes that
these potentials and parameters are relevant to the outside Vainshtein region and to the Cubic Galileon theory. The ``$1$'' denotes the 
fact that these are the first non-zero correction coming from Vainshteinian effects. We use a similar notation when we discuss the Quartic Galileon.
 
\subsection{Inside Region}

For the inside regions, it is the non-linear contribution to the scalar field kinetic term that dominates the scalar field equation.
 Given the ansatz \eqref{phianzin} and \eqref{Banz} the leading order term of the scalar equation gives an equation for 
 $U^{(in,3)}_{V_{1}}  = B^{(1,0)} /(2\sqrt{2G})$ as
\begin{align}
\ven^{i}\ven^{j} U^{(in,3)}_{V_{1}}   \ven_{i}\ven_{j} U^{(in,3)}_{V_{1}} 
-\left(\ven^{2} U^{(in,3)}_{V_{1}} \right)^{2} &=  -4\pi \rho 
\label{phin21}
\end{align}
Although it can be shown that higher orders in the expansion (\ref{phianzin}) obey linear differential equations~\cite{AvilezEtAl2015}, 
the equation for (\ref{phin21}) is non-linear and there is no known general solution. In spherical symmetry equation (\ref{phin21}) 
can be solved analytically and it has been found in~\cite{AvilezEtAl2015} that solutions agree with those presented in \cite{ChowKhoury2009}.

Once the solution for $U^{(in,3)}_{V_{1}}$ is found, the Einstein equations determine the solutions to the inside metric as~\cite{AvilezEtAl2015}
\begin{align}
h_{00}^{(2,\leq 2)} &= 2GU  +2g^{(in,3)}_{V_{1}} U^{(in,3)}_{V_{1}} 
\label{cubicinhoo} 
\\
h_{ij}^{(2,\leq 2)} &= \left[2 G U +2\gamma^{(in,3)}_{V_{1}}U^{(in,3)}_{V_{1}}\right]\gamma_{ij} 
\label{cubicinhij}
\end{align}
where for this particular theory
\begin{equation}
 \gamma^{(in,3)}_{V_{1}} = \frac{8\pi G}{\alpha_3} = \left(\frac{\Lambda}{M_{P}}\right)^{3/2}
\end{equation}
and $g^{(in,3)}_{V_{1}} = - \gamma^{(in,3)}_{V_{1}}$. Clearly, as $\alpha_3\rightarrow \infty$ the theory tends to GR.
The metric solution to $\Order(4,1)$  has been determined in \cite{McManus:2017itv}.

%%%%%%%%%%%%%%%%%%%%%%%%%%%%%%%%%%%%%%%%%%%%%%%%%%%%%%%%%%%%%%%%%%%%%%%%%%%%%%%%%%%%%%%%%%%%%%%%%%%%%%%%
 %%%%%%%%%%%%%%%%%%%%%%%%%%%%%%%%%%%%%%%%%%%%%%%%  QUARTIC 4 %%%%%%%%%%%%%%%%%%%%%%%%%%%%%%%%%%%%%%%%%%%%%%%%
\section{Quartic Galileon} 
\label{quartic}
We now discuss the case of the Quartic Galileon. The action for this theory is~\cite{Clifton:2011jh}:

\begin{widetext}
\begin{align}
S_{4}[\tilde{g},\chi] &= \int d^{4}x\sqrt{-\tilde{g}}
  \left\{
  \frac{1}{16\pi G}\tilde{R}+c_{0}\tilde{X} + \frac{1}{\Lambda^6} \tilde{X} \left[
  \left( \squareE \chi \right)^2
-\connE_\alpha \connE^\beta \chi \connE_\beta \connE^\alpha \chi
+\frac{1}{2} \RiemE \tilde{X} \right]
\right\}
 + S_{M}[g] 
   \text{.}
\label{action_quartic_galileon_Einstein}
\end{align}
\end{widetext}
where $\tilde{X}$ is defined as in (\ref{tildexdef}). Also as in the Cubic Galileon case, $\tilde{g}_{\mu\nu} = e^{2\chi/M_{P}}g_{\mu\nu}$ and for the purposes of the PPNV analysis it is useful to write the action instead in terms of $g_{\mu\nu}$ and a field $\phi\equiv e^{-2\chi/M_{P}}$. Up to a boundary term, the resulting action is found to be:
\begin{widetext}
	\begin{align}
	S'_{4}[g,\phi]  &=   \frac{1}{16\pi G} \int d^4x \volM \bigg\{ \left( \phi \RiemM + \frac{2\omega}{\phi} Y  \right)
	+   \frac{\alpha_4}{8}   \frac{Y}{\phi^5}  
	\bigg[ \left(\squareM\phi \right)^2 -   \connM_\alpha  \connM^\beta \phi  \connM_\beta  \connM^\alpha \phi+ \frac{1}{2}      \RiemM Y	+ \frac{5}{2\phi}  Y \squareM \phi
	+\frac{21}{2\phi^2} Y^2 
	\bigg] \bigg\}
\nn\\
	& + S_{M}[g]
	\label{action_quartic_galileon_Jordan_Horndeski_form}
	\end{align}
	\end{widetext}
where $Y$ and $\omega$ are defined as in (\ref{ydef}) and (\ref{omegadef}) respectively and $\alpha_{4}\equiv M_{P}^2/\Lambda^{6}$. Interestingly, the kinetic terms in (\ref{action_quartic_galileon_Jordan_Horndeski_form}) are not just those of the Quartic Galileon (with respect to $\phi$ and $Y$) but also contain a Cubic Galileon term and a  `K-essence' term proportional to $Y^{2}$.
The field equations of motion following from (\ref{action_quartic_galileon_Jordan_Horndeski_form}) are shown in appendix-\ref{eq_quartic}.

\subsection{PPNV Formalism for Quartic Galileon}
We now detail the application of the PPNV formalism to the case of the Quartic Galileon. As in the PPN case, it is assumed that - to a good approximation in regions of interest - the metric $g_{\mu\nu}$ takes the form $g_{\mu\nu} = \eta_{\mu\nu} + h_{\mu\nu}$, as in \eqref{metric_eta_h},
 where all components of $h_{\mu\nu}$ have an $N$ order of at least two. Also as in the PPN case, matter is taken to be described by a fluid and we may allocate matter density as measured by an observer momentarily freely-falling with the matter, $\rho$, as being ${\cal O}_{PPNV}(2,0)$; matter coordinate velocity $v^{i}$ as being ${\cal O}_{PPNV}(1,0)$; and co-moving matter pressure $P$ as being ${\cal O}_{PPNV}(4,0)$ (i.e. being of post-Newtonian order). Furthermore, time derivatives on a quantity are taken to raise $N$ orders by order unity whilst not changing the $V$ order.

As in the case of the Cubic Galileon, the action for the Quartic Galileon (\ref{action_quartic_galileon_Jordan_Horndeski_form}) contains a dimensionful scale - named $\alpha_{4}$ - and a vital first step is to assign a PPNV order to it. To do so, we must establish that Vainshtein screening occurs and then relate $\alpha_{4}$ to the Vainshtein radius and Schwarzschild radius. Towards this, we note that the action (\ref{action_quartic_galileon_Einstein}) for perturbations
 around Minkowski  can be written schematically as:
	
\begin{align}
S &\sim \int d^{4}x \bigg[M_{P}^{2}\sqrt{-\tilde{g}}R - (\partial \chi)^{2}
  - \frac{\partial^{6}\chi^{4}}{\Lambda^{6}}
\nn\\
&+M_{P}h\frac{\partial^{6}\chi^{3}}{\Lambda^{6}} + h_{\mu\nu}T^{\mu\nu}+\frac{\chi}{M_{p}}T\bigg]
\end{align}

From the expression above, we see that for this theory $m=6$ and $n=4$ so that $s = \frac{1}{3}$ (as in the cubic galileon)  and  $k=2$.
This  automatically gives the Vainshtein radius as $r_{V} = \frac{1}{\Lambda}\left(\frac{M}{M_{P}}\right)^{\frac{1}{3}}$ 
 Given that $\alpha_{4} = M_{P}^2/\Lambda^{6}$ then we have
\begin{align}
\alpha_{4} = \frac{r_{V}^{6}}{16\pi^2 r_{S}^2} \sim {\cal O}_{PPNV}\left(-4,-3\right) 
\end{align}
according to our order assigning prescription.

We then have the necessary ingredients to consider either a screened (inside) or un-screened (outside) region and expand the full equations according to $PPNV$ order and thus find perturbative solutions. 

\subsection{Outside Region}
Firstly we consider fields in an outside region. As in the case of the Cubic Galileon at large distances away from a Vainshtein boundary the scalar field's dynamics will be dominated by the canonical kinetic term.  Thus, in the limit of spherical symmetry $\phi \sim 1/r$, with ${\cal O}_{PPNV}(2,0)$, and we expect that corrections will go as $(r_{V}/r)$ to some positive power. Hence, we expect corrections to the dominant term to have PPNV order ${\cal O}_{PPNV}(2,V<0)$. As in the case of the Cubic Galileon, 
make the same ansatz \eqref{phianzout} as in the Cubic Galileon case.
 For the metric perturbations we similarly generalize the usual PPN order allocation to $h_{\mu\nu}$ to include a $V$ order:

\begin{align}
h_{00} &= \sum_{N=2}^{\infty}\sum_{V=0}^{-\infty}h_{00}^{(N,V)}(t,x^{i}) \label{hooout} \\
h_{0i} &= \sum_{N=3}^{\infty}\sum_{V=0}^{-\infty}h_{0i}^{(N,V)}(t,x^{i}) \label{h0iout}\\
h_{ij} &= \sum_{N=2}^{\infty}\sum_{V=0}^{-\infty}h_{ij}^{(N,V)}(t,x^{i}) \label{hijout}
\end{align}
We now focus on the solution of fields to Newtonian order ($N=2$). 
After a lengthy calculation, to this order the  Einstein equation \eqref{Quartic_galileon_Einstein_equations}, leads to the following 
 equations for the $00$ component 
\begin{align}
 -\frac{1}{2}  \grad^2 h_{00} &= \frac{4\pi G}{\phi_0^{(out)}} \rho - \frac{1}{2}  \grad^2 \phi
\label{quarti_00}
\end{align}
and for the $ij$ components
\begin{align}
&\grad_k \grad_{(i} h^k_{\;\;j)}
- \frac{1}{2} \grad^2 h_{ij}
+ \frac{1}{2} \grad_i \grad_j (h_{00} -h)
\nonumber 
\\
&
\ \ \ \
= 
 \frac{4\pi G}{\phi_0^{(out)}}  \rho \gamma_{ij}
+ \frac{1}{2} \grad^2 \varphi \gamma_{ij}
+    \grad_i \grad_j  \varphi
 \label{quartic_ij}
\end{align}
and for notational compactness it is understood that quantities in $h_{00}$, $h$ and $\vp$ are to have $N$ order 2 with a sum over $V$ orders implicit. To further simplify the equations we can impose the following gauge-fixing condition
\begin{align}
\grad_{k}h^{k}_{\ph{k}i} &= \grad_{i}\left(\frac{1}{2}h - \frac{1}{2}h_{00}+\vp\right).
\label{GF_outside}
\end{align}
Therefore \eqref{quartic_ij} turns into
\begin{align}
& - \frac{1}{2} \grad^2 h_{ij}
= 
  \frac{4\pi G}{\phi_0^{(out)}}   \rho \gamma_{ij}
+ \frac{1}{2} \grad^2 \varphi \gamma_{ij}.
 \label{quartic_ij_GF}
\end{align}
We may formally solve \eqref{quarti_00} and \eqref{quartic_ij_GF} to get
\begin{align}
 h_{00} &=  \frac{2G}{\phi_0^{(out)}} U  + \varphi 
\label{quartic_out_h_00_O2}
\\
 h_{ij} &= \left(   \frac{2G}{\phi_0^{(out)}} U  - \varphi \right) \gamma_{ij}
\label{quartic_out_h_ij_O2}
\end{align}
where $U$ is defined in equation (\ref{usol})  and so only $\varphi$ remains to be determined.

Determining $\varphi$ is achived by considering the 
  scalar field equation \eqref{Quartic_galileon_scalar_equation} which to PPN order $2$ takes the form
\begin{align}
\grad^{2}\vp &= -  \frac{8\pi G}{(3+2\omega)\phi_0^{(out)}}  \rho 
 - \frac{\alpha_{4} }{8 (3+2\omega) (\phi_0^{(out)})^2} \bigg[
\nn
\\
&
(\grad^{2} \vp)^3 - 3 \vp^{i}_{\ph{i}j} \vp^{j}_{\ph{j}i} \grad^{2} \vp +2\vp^{i}_{\ph{i}j}\vp^{j}_{\ph{j}k}\vp^{k}_{\ph{k}i}\bigg]
\label{quartic_scalar_O2}
\end{align}
where $\vp_{i} \equiv \vec{\nabla}_{i}\vp$ and $\vp_{ij}\equiv\vec{\nabla}_{i}\vec{\nabla}_{j}\vp$.

Up to this point, no considerations regarding the Vainshtein order have been made. In order to proceed further, we solve \eqref{quartic_scalar_O2}
 by expanding $\varphi$ fully as in \eqref{phianzout} and collecting Vainshtein orders.

\subsubsection{$N=2,V=0$:}
\label{quarticn2v0}
To zeroth order in $V$  the $\alpha_4$ term in the scalar field equation \eqref{quartic_scalar_O2} does not contribute and we find that
\begin{equation}
\grad^{2}\vp^{(2,0)} = -  \frac{8\pi G}{(3+2\omega)\phi_0^{(out)}}  \rho 
\end{equation}
as in standard Brans-Dicke theory. This can be formally integrated to give
\begin{equation}
\vp^{(2,0)} = -  \frac{2G}{(3+2\omega)\phi_0^{(out)}}  U   
\end{equation}
thus the metric solution to $\Order(2,0)$ is
\begin{align}
 h_{00}^{(2,0)} &=  2 G_C U
\\
 h_{ij}^{(2,0)} &= 2 \gamma G_C U \gamma_{ij}
\end{align}
where $G_C$ and $\gamma$ are  just as in Brans-Dicke and the Cubic Galileon theories given by \eqref{G_C} and \eqref{gamma_PPN}  respectively.
 Recall that in the limit that non-linear kinetic terms may be ignored, the action (\ref{action_quartic_galileon_Jordan_Horndeski_form})
 corresponds to that of Brans-Dicke theory. Indeed  the solution to this order is identical to the 
 Newtonian limit of that theory  \cite{Brans1961}.

\subsubsection{$N=2,V=-3$:} 
%To order $N=2,V=-4$ we have 
The presence of $\alpha_4$ in the scalar equation \eqref{quartic_scalar_O2} implies that the first 
  non-trivial order for $V$ is $V=-3$. The scalar field equation takes the form:
\begin{align}
 \grad^{2}\vp^{(2,-3)}
&= -\frac{\alpha_{4}}{8(3+2\omega) (\phi_0^{(out)})^2} \bigg[(\grad^{2}\vp^{(2,0)})^3
\nn\\
& - 3 (\vp^{(2,0)})^{i}_{\ph{i}j} (\vp^{(2,0)})^{j}_{\ph{j}i} \grad^{2} (\vp^{(2,0)})
\nn\\
&+2(\vp^{(2,0)})^{i}_{\ph{i}j}(\vp^{(2,0)})^{j}_{\ph{j}k}(\vp^{(2,0)})^{k}_{\ph{k}i}\bigg] 
\label{phi2min4eq}
\end{align}
Defining the potential $U^{(out,4)}_{V_{1}}$ via
\begin{align}
\vp^{(2,-3)} &=  \frac{\alpha_{4} G_C^3 }{32\pi (3+2\omega) (2+\omega)^3(\phi_0^{(out)})^2 }U^{(out,4)}_{V_{1}} 
\end{align}
we may formally solve \eqref{phi2min4eq} to obtain
\begin{align}
U^{(out,4)}_{V_{1}} \equiv& \int \frac{d^{3}\vec{x}'}{|\vec{x}-\vec{x}'|}
  \bigg\{  4\pi \rho \left[ - (4\pi \rho)^{2} + 3 U^{ij}U_{ij}\right] 
 \nonumber
\\
&
+2U^{i}_{\ph{i}j} U^{j}_{\ph{j}k}U^{k}_{\ph{k}i}\bigg\}
\label{uv41}
\end{align}
where $U_{ij} = \grad_i \grad_j U$.

Therefore, summing up the $\Order(2,0)$ and $\Order(2,-3)$ contributions in \eqref{quartic_out_h_00_O2} and \eqref{quartic_out_h_ij_O2}
we arrive at the metric solution
\begin{align}
 h_{00} &=  2G_C U   +  2 g^{(out,4)}_{V_{1}}  \left(\frac{M_p}{H_0}\right)^6 G_C^5  U^{(out,4)}_{V_{1}} 
 \label{quarticouthoo} 
\\
 h_{ij} &=  \left[ 2G_C U   +  2 \gamma^{(out,4)}_{V_{1}}  \left(\frac{M_p}{H_0}\right)^6 G_C^5  U^{(out,4)}_{V_{1}} \right]\gamma_{ij}
\label{quarticouthij}
\end{align}
where
%\begin{align}
% \gamma^{(out,4)}_{V_{1}} &\equiv \frac{1}{(4+2\omega)^{4}} M_p^4  \alpha_{4}^{2} \left(\frac{H_{0}}{M_{P}}\right)^{4} 
%\end{align}
\begin{align}
 \gamma^{(out,4)}_{V_{1}} &\equiv - \frac{\pi}{4} \frac{2\omega+3}{(2+\omega)^5} \left( \frac{H_0}{\Lambda}\right)^6
 = - \frac{\pi}{4} \frac{2\omega+3}{(2+\omega)^5}\frac{H_0^6}{M_p^2} \alpha_4
\label{gv41}
\end{align}
and $g^{(out,4)}_{V_{1}} = - \gamma^{(out,4)}_{V_{1}}$.
We see then that for the Quartic Galileon, there exists - to Newtonian order - a new PPNV potential (\ref{uv41}) with accompanying dimensionless 
PPNV parameters $g^{(out,4)}_{V_{1}}$ and $\gamma^{(out,4)}_{V_{1}}$. As in the case of the Cubic Galileon PPNV parameter $\gamma^{(out,3)}_{V_{1}}$, 
factors of $H_{0}$ have been included so that for values of $\Lambda$ such that the scalar field has a role to play in late time cosmology, 
deviations of the value $\gamma^{(out,4)}_{V_{1}}$ from unity are largely controlled by the value of $\omega$.

\subsection{Inside region}
Now we consider the behavior of fields in the inside region to Newtonian $N=2$ order.
Applying the dualization strategy as described above in section \ref{cubic} leads to
\begin{align}
\nabla_{\mu}\varphi  = \alpha_{4}^{-1/3} \nabla_{\mu}B
 \label{phitob}
\end{align}
For reference recall that we have that $\alpha_{4}^{-1/3}$ is of 
 PPNV order 
\begin{equation}
 \alpha_{4}^{-1/3} \sim {\cal O}_{PPNV}(\frac{4}{3},1).
\end{equation}
while $B$ has PPN order $2/3$.

The dualized field equations are displayed in the appendix, \ref{app_dual}, and 
under (\ref{phitob}), the $N=2$ order Einstein equations \eqref{Quartic_galileon_dual_Einstein_equations}  turn into
\begin{align}
&
-\frac{1}{2}\grad^{2} h_{00} =  4\pi G\rho -  \frac{1}{2\alpha_{4}^{1/3}}   \grad^2 B   
\end{align}
and
\begin{align}
&
  \grad_k \grad_{(i} h^k_{\;\;j)}
- \frac{1}{2} \grad^2 h_{ij}
+ \frac{1}{2} \grad_i \grad_j (h_{00} -h)
\nonumber
\\
&=  4\pi G\rho \gamma_{ij}
 +  \frac{1}{\alpha_{4}^{1/3}} \left(  \grad_i \grad_j B +    \frac{1}{2}   \grad^2 B   \gamma_{ij} \right).
\label{Ein_quart_ij}
\end{align}
Imposing the gauge-fixing condition 
\begin{equation}
\grad_{k}h^{k}_{\ph{k}i} = \grad_{i}\left(\frac{1}{2}h - \frac{1}{2}h_{00} + \alpha_{4}^{-1/3} B\right).
\label{GF_inside}
\end{equation}
brings  \eqref{Ein_quart_ij} into
\begin{align}
& - \frac{1}{2} \grad^2 h_{ij} = \left[  4\pi G\rho +  \frac{1}{2\alpha_{4}^{1/3}}    \grad^2 B \right]  \gamma_{ij} 
\end{align}
thus the formal metric solution is
\begin{align}
h_{00} &= 2 G U + \alpha_{4}^{-1/3} B
\label{quartic_in_h00_O2}
\\
h_{ij} &= \left(2 G U - \alpha_{4}^{-1/3} B \right)\gamma_{ij}
\label{quartic_in_hij_O2}
\end{align}
Note that although the gauge-fixing condition \eqref{GF_inside} is the same as \eqref{GF_outside}, matching the inside to the outside solutions is impossible as any solution is inhertently non-perturbative at the Vainshtein radius $r_V$.
In order to determine $B$ we use the scalar equation \eqref{Quartic_galileon_dual_scalar_equation} which
 to this order is
\begin{align}
& (\grad^{2} B)^3 - 3 B^{i}_{\ph{i}j} B^{j}_{\ph{j}i} \grad^{2} B +2B^{i}_{\ph{i}j}B^{j}_{\ph{j}k}B^{k}_{\ph{k}i}
\nn
\\
&
 \ \ \ \
+ 8\alpha_{4}^{-1/3}(3+2\omega) \grad^{2}B
= -  64\pi G \rho 
\label{quartic_in_scalar_O2}
\end{align}
Now we make the following ansatz for our fields 
\begin{align}
B^{(2/3)}  &=  \sum_{V=0}^{\infty}B^{(2/3,V)}
\label{quartic_inside_scalar_series}
\end{align}

and determine the solutions order-by-order in $V$.

\subsubsection{$N=2,V=0$}
The leading order contribution from $B$ is $B^{(2/3,0)}$ so that owing to the $\alpha_4^{-1/3}$ term in \eqref{quartic_in_h00_O2} and \eqref{quartic_in_hij_O2} 
it drops out and the metric solution to this order is exactly as in GR. That is, to order $(2,0)$ the metric components are
$h_{00} = 2 G U$ and $h_{ij} = 2G U \gamma_{ij}$.
This is an explicit realization of the Vainshtein mechanism which is found to be active in this theory, just as in the case of the Cubic Galileon.

\subsubsection{$N=2,V=1$}
To determine the next order contribution to  \eqref{quartic_in_h00_O2} and \eqref{quartic_in_hij_O2}  it suffices to determine $B^{(2/3,0)}$ using the 
scalar field equation \eqref{quartic_in_scalar_O2} to this order. 
We let, for notational compactness, $B^{(2/3,0)}  \equiv -2(2G)^{1/3}  U^{(in,4)}_{V_1}  $ so that 
introducing the matrix notation ${\mathbf B} \leftrightarrow \grad^i \grad_j   U^{(in,4)}_{V_1}$
the scalar field equation takes the form
\begin{align}
& (\Tr{\mathbf B})^3 - 3  \Tr( {\mathbf B}^2 ) (\Tr{\mathbf B}) + 2 \Tr({\mathbf B}^3) =   4\pi  \rho 
\label{quartic_in_scalar_LO}
\end{align}
%
%the scalar field equation takes the form:
%
%\begin{align}
%(\grad^{2} {\cal B})^3 - 3 {\cal B}^{i}_{\ph{i}j} {\cal B}^{j}_{\ph{j}i} \grad^{2} {\cal B} +2{\cal B}^{i}_{\ph{i}j}{\cal B}^{j}_{\ph{j}k}{\cal B}^{k}_{\ph{k}i}
%&= 4\pi  \rho 
%\label{calbeq}
%\end{align}
%
%This explains the slightly unusual $N$ order of the field $B^{(2/3,0)}$ i.e. the field appears cubically to create an overall $N=2$ contribution. 
We see then that as in the case of the Cubic Galileon, the leading contribution to the scalar field equation is a non-linear partial differential equation. 

To our best of knowledge, just like in the case of the Cubic Galileon, it is impossible to write $U^{(in,4)}_{V_1}$ 
in integral form, except in the case of spherical
symmetry which we present further below. However, \eqref{quartic_in_scalar_LO} may in principle  be solved numerically, or using perturbation techniques around spherical symmetry. Assuming that we do have $ U^{(in,4)}_{V_1} $ at hand from such procedures, the metric solution to this order is
\begin{align}
h^{(2,\leq 4/3)}_{00} &= 2 G U + 2 g^{(in,4)}_{V_{1}}  G^{-1/3}  U^{(in,4)}_{V_1} 
\label{quartic_in_h00}
\\
h^{(2,\leq 4/3)}_{ij} &= \left(2 G U + 2 \gamma^{(in,4)}_{V_{1}}  G^{-1/3}   U^{(in,4)}_{V_1}  \right)\gamma_{ij}
\label{quartic_in_hij}
\end{align}
where 
\begin{align}
\gamma^{(in,4)}_{V_{1}} &=  (32\pi^2)^{-1/3}  \left(\frac{\Lambda}{M_p}\right)^{2} 
\label{gamma4in}
%\gamma_{V_{4(2)}} &= \frac{1}{2}\frac{\Lambda^{2}}{(M_{P}H^{2}_{0})^{2/3}} 
\end{align}
and $ g^{(in,4)}_{V_{1}}  = - \gamma^{(in,4)}_{V_{1}}$ are two PPNV parameters. Note that since the potential $U^{(in,4)}_{V_1}$ is different to
$U^{(in,3)}_{V_1}$,  these parameters are distinct from the case of the Cubic Galileon.
%
%where we have introduced the constant $\gamma_{V_{4(2)}} $. In summary we have then that
%
%\begin{align}
%h_{00}^{(2,\leq 4/3)} &= 2G_{N}U +2\gamma_{V_{4(2)}}H_{0}^{3/4}G^{(1/3)}{\cal B} \\
%h_{ij}^{(2,\leq 4/3)} &=  \bigg[2G_{N} U -2\gamma_{V_{4(2)}}H_{0}^{3/4}G^{(1/3)}{\cal B}  \bigg] \gamma_{ij}
%\end{align}
%
%where $G_{N} = G/\phi_{0}^{(in)}$ is the Newton's constant that would be measured in the solar system and ${\cal B}$ is a solution to  (\ref{calbeq}).

%%%%%%%%%%%%%%%%%%%%%%%%%%%%%%%%% SPHERICAL QUARTIC %%%%%%%%%%%%%%%%%%%%%%%%%%%%%%%%%%%%%%%%%%%%%%%%%%%%\

\subsubsection{$N=2,V=2$}
We may continue our iteration to determine the next  correction to $h_{00}$, i.e. to Vainshteinian order $V=2$. This is achieved by 
expanding the scalar equation \eqref{quartic_in_scalar_O2}  to order $V=1$ using \eqref{quartic_inside_scalar_series}  and collecting terms.
Setting  $B^{(2/3,1)}  \equiv  -\frac{4(3+2\omega)}{3 \alpha^{2/3} (2G)^{1/3} }  U^{(in,4)}_{V_2} $ and introducing
the matrix notation ${\mathbf C} \leftrightarrow  \grad^i \grad_j   U^{(in,4)}_{V_2}$ this leads to the linear equation
\begin{align}
& 
 \left[  (\Tr{\mathbf B})^2 -   (\Tr{\mathbf B}^2)  \right] \Tr{\mathbf C} 
- 2 (\Tr{\mathbf B})  \Tr({\mathbf B} {\mathbf C}) 
+ 2\Tr({\mathbf B}^2 {\mathbf C})
\nn
\\
&
 \ \ \ \
+ \Tr{\mathbf B}
=  0
\label{quartic_in_scalar_Next_V}
\end{align}
where at this stage the solution for $ U^{(in,4)}_{V_1}$ is assumed 
as determined by the previous step.

One may continue the iteration to higher Vainshteinian orders as necessary, each time resulting to a linear equation for the next order where the 
previous orders are used as sources.

\subsection{Spherical symmetry}
The solutions found for contributions to the scalar field and metric tensor in the outside region are rather complicated whilst in the inside region we are faced with equations that have no known general solution. To aid intuition, we can restrict ourselves to spherical symmetry; this will enable us to obtain some simple solutions for quantities of interest. To this end, we will assume that the matter source mass is a spherically symmetric mass $M$ and uniform density between $r=0$ and $r=r_{*}$ (and zero for $r>r_{*}$). Firstly we consider the outside (unscreened) region. 
Let us also note that exterior spherically symmetric solutions inside the Vainshtein radius have been found earlier in \cite{Burrage:2010rs}
while both interior and exterior solutions inside and outside the Vainshtein radius have been determined 
in~\cite{Bloomfield:2014zfa}.  The purpose of this subsection is to firstly serve as a consistency check with these known results and also
to extend them, as we discuss below, to include the next Vainshteinian corrections not included in \cite{Bloomfield:2014zfa}

\subsubsection{The outside region}
We assume a spherically symmetric source of mass $M$ and radius $r_*$ so that
\begin{align}
\rho(r) = \frac{3M}{4\pi r_{*}^{3}}\Theta(r_{*}-r)
\end{align}
where  $\Theta(x)$ is the Heaviside step function ($\Theta(x) = 1$ for $x\geq 0$, and $\Theta(x)=0$ for $x<0$).

We then abreviate the matrix $U^i_{\;\;j}$ in \eqref{uv41} as ${\bf U} =
 \left[ U_{rr}   -  \frac{1}{r} U_r \right] \hat{r} \otimes  \hat{r} +  \frac{1}{r} U_r {\mathbb I}  $
where $U_r \equiv dU/dr$, the vector $\hat{r} \leftrightarrow \grad_i r$ is unit, and ${\mathbb I} \leftrightarrow \delta^i_{\;\;j}$ is the unit matrix.
Meanwhile, the Newtonian potential is given by 
\begin{equation}
U = \frac{M(3r_*^2-r^2)}{2r_*^3}\Theta(r_*-r) + \frac{M}{r} \Theta(r-r_*)
\end{equation}
so that \eqref{uv41} evaluates to
\begin{align}
U^{(out,4)}_{V_{1}} \equiv&  \frac{4\pi M^3}{7}\left[ \frac{ 7r^2 - 9r_*^2 }{r_*^9} \Theta(r_*-r) -  \frac{2}{r^7} \Theta(r_*-r) \right]
%\\
%\frac{dU}{dr} \equiv&  8\pi M^3\left[ \frac{r}{r_*^9} \Theta(r_*-r) +  \frac{1}{r^8} \Theta(r_*-r) \right]
%\\
%\grad^2 U\equiv&  -4\pi\left[ -\frac{6M^3}{r_*^9} \Theta(r_*-r) +  \frac{12M^3}{r^9} \Theta(r_*-r) \right]
%\\
%\grad^2 U\equiv&  -4\pi S
\end{align}
From this point onwards let us consider only the exterior solution to the source, i.e. $r> r_*$ as is appropriate for most physical systems of interest.
In that case, the metric solution is
\begin{align}
 h_{00} &= \frac{2G_C M}{r} \left[1  - \frac{1}{14(8\pi)^2 (2+\omega)(3+2\omega)^3 \phi_0^4 }  \frac{r_V^6}{r^6} \right]
\label{h00out}
\\
 h_{ij} &= \frac{2G_C M}{r} \left[1  + \frac{1}{14(8\pi)^2(2+\omega)(3+2\omega)^3 \phi_0^4   } \frac{r_V^6}{r^6}  \right]\gamma_{ij}
\end{align}

It is instructive to compare the size of the leading Vainshteinian contribution to the metric potentials above, at the Vainshtein radius $r_{V}$; 
their comparative effect will only decrease at larger radii. 
It is found that at $r_{V}$, the ratio of first and second terms in (\ref{h00out}) is of the order $1/\big( (\phi_0^{(out)})^{4}(3+2\omega)^{3}(1+\omega)\big)$ 
(where we have assumed that $\Lambda^{2}= {\cal O}((M_{p}H_{0}^{2})^{2/3})$. If it is the case that cosmologically the theory (\ref{action_quartic_galileon_Jordan_Horndeski_form}) is essentially that of Brans-Dicke theory (i.e. the influence of non-linear terms in the action have negligible effect) then we may use cosmological constraints on that theory restrict $\omega \gtrsim 1000$ \cite{AvilezSkordis2014} and so the force at $r_{V}$ due to first and second terms (\ref{h00out}) is of order $\omega^{-4}\sim 10^{-12}$, with this ratio only decreasing for $r>r_{V}$. Therefore the modification to Newtonian gravity in the outside region is small.

\subsubsection{The inside region}
In spherical symmetry, the equation (\ref{quartic_in_scalar_LO}) simplifies significantly, taking the non-linear ODE form:
\begin{align}
\frac{6}{r^{2}} \left(\frac{d}{dr}  U^{(in,4)}_{V_1}\right)^{2} \left(\frac{d^2}{dr^2}   U^{(in,4)}_{V_1} \right) & = 4 \pi  \rho(r) 
\label{beq2}
\end{align}

%\begin{align}
%\grad_i r &= r_i/r  = \hat{r}_i
%\\
%\grad_i \hat{r}_j &= (\delta_{ij}  -  \hat{r}_i  \hat{r}_j ) /r
%\\
% \grad_i U &= U_r \hat{r}_i
%\\
% \grad_{ij} U &= [U_{rr}   -  \frac{1}{r} U_r]\hat{r}_i    \hat{r}_j +  \frac{1}{r} U_r \delta_{ij} 
%\\
%X &= U_{rr}   -  \frac{1}{r} U_r
%\\
%Z &= U_{rr}   +  \frac{1}{r} U_r
%\\
%Y &=  U_{rr}   +  \frac{2}{r} U_r 
%\\
% B &=   X \hat{r} \otimes  \hat{r}  +  \frac{1}{r} U_r I
%\\
% \Tr(B) &=   Y
%\\
% B^2&=  XZ \hat{r} \otimes  \hat{r}    +  (\frac{1}{r} U_r)^2 I  
%\\
%\Tr B^2 &=    XZ     + 3  (\frac{1}{r} U_r)^2  
%\\
% B^3&=  
%  X \left[ XZ +  (\frac{1}{r} U_r)^2  +   Z \frac{1}{r} U_r  \right] \hat{r} \otimes  \hat{r}    +  (\frac{1}{r} U_r)^3 I  
%\\
%\Tr(B^3) &=   X \left[ XZ +  \frac{1}{r} U_r Y  \right] + 3 (\frac{1}{r} U_r)^3
%\\
%&
%  U_{rr}\left[   4  \frac{1}{r} U_r - 2  (U_{rr} )^2    -2  (\frac{1}{r} U_r)^2    \right]
%\nonumber
%\\
%& 
% 6 (\frac{1}{r} U_r)^2   U_{rr} = -  4\pi  \rho 
%\end{align}

%\begin{align}
%& (\Tr{\mathbf B})^3 - 3  \Tr( {\mathbf B}^2 ) (\Tr{\mathbf B}) + 2 \Tr({\mathbf B}^3) = -  4\pi  \rho 
%\label{quartic_in_scalar_LO}
% 6 (\frac{1}{r} U_r)^2   U_{rr} =   4\pi  \rho 
%\end{align}

%
We assume a spherically symmetric source of mass $M$ and radius $r_*$ so that
\begin{align}
\rho(r) = \frac{3M}{4\pi r_{*}^{3}}\Theta(r_{*}-r)
\end{align}
where  $\Theta(x)$ is the Heaviside step function ($\Theta(x) = 1$ for $x\geq 0$, and $\Theta(x)=0$ for $x < 0$).
 Hence we can write (\ref{beq2}) as§
\begin{align}§
 \frac{d}{dr} \left[ \left(\frac{d}{dr}  U^{(in,4)}_{V_1}\right)^{3} \right] & =   \frac{3M r^2}{2 r_{*}^{3}}\Theta(r_{*}-r)
\end{align}
Integrating twice leads to
%\begin{align}
% \left(\frac{d}{dr}  U^{(in,4)}_{V_1}\right)^{3}  & =  \frac{M}{2} \left[\frac{r^3}{r_{*}^{3}} \Theta(r_{*}-r) + \Theta(r- r_{*})  \right]
%\end{align}
%which is further integrated to
\begin{align}
 U^{(in,4)}_{V_1} &= 
 \frac{M^{1/3}}{2^{1/3}}  \left( - \frac{3}{2}  r_*  + \frac{r^2}{2r_{*}} \right) \Theta(r_{*}-r) 
\nonumber 
\\
&
  + \frac{M^{1/3}}{2^{1/3}} \left(r - \frac{r_*}{2}\right)\Theta(r- r_{*})  
\label{binsol}
\end{align}
where the meaning of the chosen integration constant is clarified further below.

Interestingly, for interior solutions, i.e. $r<r_*$ we find that $U^{(in,4)}_{V_1}  \propto (3r_*^2 - r^2)$ which is of the same $r$ dependence as the interior
solution for $U(r)$. More specifically, the solution inside the source is
%\begin{widetext}
%\begin{align}
%\Lambda^2 &= \frac{1}{r_V^2} (8\pi G)^{1/3} M^{2/3}
%\\
%2G U &= \frac{3GM}{r_*} - \frac{GM}{r_*^3} r^2
%\\
%U^{(in,4)}_{V_1} &= I^{(in,4)}_{V_1} + \frac{M^{1/3}}{2^{1/3}}\frac{r^2}{2r_{*}}
%\\
%\gamma^{(in,4)}_{V_{1}} &=  2^{-5/3} \pi^{-2/3} \frac{1}{r_V^2} (8\pi G)^{4/3} M^{2/3} 
%\\
%X^{(in)} &= \gamma^{(in,4)}_{V_{1}} G^{-1/3} U^{(in,4)}_{V_1}  
%= 2^{7/3} \pi^{2/3} \frac{1}{r_V^2}  M^{2/3} G I^{(in,4)}_{V_1} 
%+ 2 \pi^{2/3} \frac{r_*^2}{r_V^2} \frac{GM}{r_{*}^3} r^2
%\\
%X^{(in)} &= \gamma^{(in,4)}_{V_{1}} G^{-1/3} U^{(in,4)}_{V_1}  
%=
% - 6 \pi^{2/3} \frac{G M r_*}{r_V^2}  +  2 \pi^{2/3} \frac{r_*^2}{r_V^2} \frac{GM}{r_{*}^3} r^2
%\end{align}
\begin{align}
h^{(2,\leq 4/3)}_{00}|_{r<r_*}&=2 G U^{{\rm(inter,4)}}_{V_1}
\label{quartic_interior_00}
\\
h^{(2,\leq 4/3)}_{ij}|_{r<r_*} &= 2\gamma^{{\rm(inter,4)}}_{V_1} G U^{{\rm(inter,4)}}_{V_1} \gamma_{ij}
\label{quartic_interior_ij}
\end{align}
where the interior potential 
\begin{equation}
U^{{\rm(inter,4)}}_{V_1}  = \frac{3M_{{\rm eff}}}{2r_*} - \frac{M_{{\rm eff}}}{2r_*^3} r^2 
\end{equation}
is identical to the one in GR but with the mass renormalized to $M_{{\rm eff}} = M\left[1 +4\pi^{2/3} (r_*/r_V)^2 \right]$. 
The arbitrary integration constant in \eqref{binsol}  was chosen so that this identification was possible.
However, even more interestingly, the PPN $\gamma$ parameter for this solution is not unity as it should have been for GR, but rather it is
\begin{equation}
\gamma^{{\rm(inter,4)}}_{V_1} = \frac{1 - 4 \pi^{2/3} \frac{r_*^2}{r_V^2} }{ 1 +4\pi^{2/3} \frac{r_*^2}{r_V^2}  }
\end{equation}
so that the presence of the Galileon inside the source breaks the Vainshtein mechanism and introduces an effective $\gamma$ parameter.
This is of academic interest only, of course, as for usual cases of interest $r_V \gg r_*$ so that these corrections are tiny 
and the spacetime inside the source can be taken to be identical to the one in GR.

Turning now to exterior solutions, i.e. $r>r_*$ we find that the metric solution is
\begin{align}
h^{(2,\leq 4/3)}_{00}|_{r>r_*} &=    2GM \left[ \frac{1}{r} +   4 \pi^{2/3} \frac{1}{r_V^2} (  2 r_* - r) \right]
\\
h^{(2,\leq 4/3)}_{ij}|_{r>r_*} &=    2GM  \left[ \frac{1}{r} -  4 \pi^{2/3} \frac{1}{r_V^2} ( 2  r_* - r ) \right]\gamma_{ij}
\end{align}
Thus we see that in the inside Vainshtein region in spherical symmetry at Newtonian order there is a correction to the Newtonian potential 
that is proportional to the coordinate $r$. As such, the theory produces an additional, constant force exterior to a spherically symmetric body.

The presence of the constant part in the above solutions is an artifact of having chosen the arbitrary integration constant in \eqref{binsol} in order to obtain
the specific form for the interior solution \eqref{quartic_interior_00} and \eqref{quartic_interior_ij}. Had we chosen it 
such that $U^{(in,4)}_{V_1}\rightarrow 0$ as $r\rightarrow 0$ then the constant in the above solution would not be there. Naturally, such a constant does not have any physical significance.

Although the  extra force produced by the correction to the Newtonian potential in the inside region is very small, by looking at systems like binary pulsars with observations integrated over a long period of time, this effect may still be observable~\cite{Shao2012, Shao2013, Foster2007,Yagi2013}.

%\CSC{As an aid to intuition about how the Cubic Galileon PPNV formalism compared to that of the Quartic Galileon, Figure \ref{orders4} shows the necessary PPNV order that one must go to achieve a certain amount of accuracy to describe the scalar field profile due to a black hole source. It can be seen that as compared to Figure \ref{orders}, the required order differs between Cubic and Quartic cases. This is a matter of convention and ultimately arises from ...(to be finished)}
%
%\begin{figure}[h!]
%	\center
%	\epsfig{file=figure2.pdf,width=9.cm}
%	\caption{Schematic illustration for the Quartic Galileon case of the estimated PPNV order in the scalar field $\phi$ that one should go to, to achieve a certain standard of approximation to the results from solving the full equations exterior to a black hole with Schwarzschild radius $r_{S}$ and Vainshtain radius $r_{V}$. The non-continuity of the $V$ Order curve either side of $r_{V}$ illustrates the need for use of a dual formulation of the theory on the inside regions}
%	\label{orders4}
%\end{figure}
%

For the sake of completeness, let us calculate the next Vainshteinian correction to the above solutions in the case of spherical symmetry.
Adapting \eqref{quartic_in_scalar_Next_V} to spherical symmetry gives
\begin{align}
U^{(in,4)}_{V_2} = -\frac{1}{2} \int \frac{r^2}{dU^{(in,4)}_{V_1} / dr  } dr.
\end{align}
Plugging in the determined solutions for $U^{(in,4)}_{V_1}$ and integrating gives
\begin{align}
  U^{(in,4)}_{V_2} \propto r^2 \qquad \mbox{for} \quad r\le r_*
\\
  U^{(in,4)}_{V_2} \propto r^3  \qquad \mbox{for} \quad r\ge r_*
\end{align}
Interestingly, the interior solutions do not acquire any further powers of $r$ but retain their form \eqref{quartic_interior_00} with the constants
appearing in the solution appropriately renormalised while the next correction to the exterior solution goes as $r^3$ rather as $r^2$.

\section{Quintic Galileon} 
\label{quintic}
Finally we discuss the case of the Quintic Galileon. The action for this theory is:

\begin{widetext}
	\begin{align}
	\tilde{S}_{5}[\tilde{g},\chi]&= \int d^{4}x \sqrt{-\tilde{g}}\bigg\{
\frac{1}{16\pi G}\tilde{R}+c_{0}\tilde{X}
\nn
\\
& -\frac{1}{\Lambda^{9}}\tilde{X}\left[(\tilde\Box\chi)^{3}-3\tilde\Box\chi\tn_{\mu}\tn_{\nu}\chi\tn^{\mu}\tn^{\nu}\chi+2\tn_{\mu}\tn_{\nu}\chi\tn^{\nu}\tn^{\alpha}\chi\tn_{\alpha}\tn^{\mu}\chi-3\tx\tilde{G}^{\mu\nu}\tn_{\mu}\tn_{\nu}\chi\right]
\bigg\}
+S_{M}[g] 
\label{quinein}
	\end{align}
\end{widetext}
As in the case of the Cubic Galileon and Quartic Galileon theories, it is useful to write the theory in terms of fields $g_{\mu\nu}$ and $\phi$. This can be done, yielding, up to boundary terms, the following action:

\begin{widetext}
	\begin{align}
	S_{5}[g,\phi] &=\frac{1}{16\pi G}\int d^{4}x\sqrt{-g}\bigg\{\phi R
	+\frac{2\omega}{\phi}Y
	+\frac{\alpha_{5}}{16\phi^{7}}Y\left[
 (\Box\phi)^{3}-3\Box\phi\phi_{\mu\nu}\phi^{\mu\nu}+2\phi_{\mu\nu}\phi^{\nu\alpha}\phi_{\alpha\ph{\mu}}^{\ph{\gamma}\mu}-3YG^{\mu\nu}\phi_{\mu\nu}\right]
\nn
\\
&  
+\frac{9\alpha_{5}}{16\phi^{8}}Y^{2}\left[  (\Box\phi)^{2}-\phi_{\mu\nu}\phi^{\mu\nu}+\frac{1}{3}YR\right]
  +\frac{63}{16}\frac{\alpha_{5}}{\phi^{9}} Y^{3}\Box\phi +\frac{39}{4}\frac{\alpha_{5}}{\phi^{10}}Y^{4} 
\bigg\}
+ S_{M}[g]
 \label{quintjor}
	\end{align}
\end{widetext}
where $Y$ and $\omega$ are defined as in (\ref{ydef}) and (\ref{omegadef}) respectively and $\alpha_{5}\equiv M_{P}^3/\Lambda^{9}$.
Interestingly, the kinetic terms in (\ref{quintjor}) are not just those of the Quintic Galileon (with respect to $\phi$ and $Y$)
 but also contain a Cubic and a Quartic Galileon term and a  `K-essence' term proportional to $Y^{4}$.  
The field equations of motion following from (\ref{action_quartic_galileon_Jordan_Horndeski_form}) are shown in appendix-\ref{eq_quintic}.
%As in the case of the Quartic Galileon, the full equations of motion following from (\ref{quintjor}) may be read off from the results contained in Appendices A and B of (\cite{Kobayashi:2011nu}).

%%%%%%%%%%%%%%%%%%%%%%%%%%%%%%%%%%%%%%%%%% PPNV QUINTIC %%%%%%%%%%%%%%%%%%%%%%%%%%%%%%%%%%%%%%%%
\subsection{PPNV Formalism for Quintic Galileon}
In this part we detail the application of the PPNV formalism to the case of the Quintic Galileon. As in the case of $\alpha_{3}$ and $\alpha_{4}$ 
for the Cubic Galileon and Quartic Galileon, it is important to assign a PPNV order to the dimensionful constant $\alpha_{5}$ that appears in the 
Quintic Galileon action (\ref{quintjor}).  The schematic form of 
the action for the Quintic Galileon for perturbations around a Minkowski space background, is as follows:

	\begin{align}
	S &\sim \int d^{4}x \bigg[M_{P}^{2}\sqrt{-\tilde{g}}R - (\partial \chi)^{2}+ \frac{\partial^{8}\chi^{5}}{\Lambda^{9}}\nn\\
	&+M_{P}h\frac{\partial^{8}\chi^{3}}{\Lambda^{6}} + h_{\mu\nu}T^{\mu\nu}+\frac{\chi}{M_{p}}T\bigg]
	\end{align}
so that for this theory $m=8$ and $n=5$  giving $k=9/4$ and $s=1/3$. Thus, once again the
 Vainshtein radius is $r_V = \frac{1}{\Lambda}\left(\frac{M}{M_{P}}\right)^{\frac{1}{3}}$, 
 Given that $\alpha_{5} = M_{P}^3/\Lambda^{9}$ then we have
\begin{align}
\alpha_{5} = \frac{r_{V}^{9}}{64\pi^3 r_{S}^3} \sim {\cal O}_{PPNV}\left(-6,-4\right) 
\end{align}
according to our order assigning prescription.

%%%%%%%%%%%%%%%%%%%%%%%%%%%%%%%%%%%%%%%%%%%%%%%%% OUTSIDE  QUINTIC %%%%%%%%%%%%%%%%%%%%%%%%%%%%%%%

\subsection{Non-Existence of Vainshtein Mechanism to order $N=2$}

Firstly we consider fields in an outside region. We proceed in a similar way to as in the Quartic Galileon case 
and so assign the same PPNV orders to matter quantities and make the same ansatz for scalar field $\phi$ and metric tensor perturbation $h_{\mu\nu}$ as 
\eqref{phianzout}, \eqref{hooout}, \eqref{h0iout} and \eqref{hijout}. To Newtonian order the scalar field equation \eqref{eq_quintic_ordinary_scalar} takes the form
\begin{align}
 - (2\omega  +3) \grad^2 \varphi &= \frac{8\pi G}{\phi_0} \rho
+  \frac{\alpha_5}{16} \frac{1}{\phi_0^3}
  \bigg[ (\Tr \Mb)^4 
 + 3 ( \Tr \Mb^2)^2
\nn
\\
&
 - 6  (\Tr \Mb)^2  \Tr(\Mb^2)
 +  8(\Tr \Mb) \Tr(\Mb^3)
\nn
\\
&
 - 6 \Tr(\Mb^4) \bigg] 
\label{quintnewton}
\end{align}
where we introduced the matrix $\Mb \leftrightarrow \phi^i_{\;\;j}$.

Equation \eqref{quintnewton} looks rather similar in structure to (\ref{phi2min4eq}) in the case of the Quartic Galileon wherein there $\vp^{(2,-4)}$ 
in the outside region was sourced by non-linear terms in $\vp^{(2,0)}$. 
However, there is an important difference:  the term in square brackets in \eqref{quintnewton} 
is \emph{identically zero}. This may be understood as follows: the matrix $\Mb$ is a pullback to 
spatial components of the $4\times4$ matrix $\phi_{\mu\nu}$ and the term is proportional to $\mathrm{det}(\phi_{\mu\nu})$; to this 
PPNV order, only $ij$ terms contribute and so $\phi_{\mu\nu}$ has vanishing determinant.
Thus, the scalar field equation \eqref{quintnewton} is as in Brans-Dicke theory and a straight forward inspection of the Einstein equations
\eqref{quintnewton} shows that  to Newtonian order they are also as in  Brans-Dicke.

%\begin{align}
%\grad^{2}\vp^{(2,-6)} &= 0  
%\end{align}
%
%Indeed, to order $N=2$ (without decomposing $\vp$ by $V$ order), the scalar field equation for the Quintic Galileon is:
%
%\begin{align}
%- \grad^{2}\phi &=\frac{8\pi G}{(3+2\omega)}\rho \nn\\
%& + \frac{\alpha_{5}^{3}}{16 (\phi_0^{(out)})^{6}}\frac{1}{(3+2\omega)}\bigg[
%(\grad^{2} \phi)^4  - 6 \phi^{ij}\phi_{ij}(\grad^{2} \phi)^2 \nn\\
%& + 3(\phi^{ij}\phi_{ij})^{2}
%+  8\grad^{2}\phi \phi^{ik}\phi^{j}_{\ph{j}k}\phi_{ij} 
%- 6\phi^{ik}\phi^{j}_{\ph{j}k}\phi_{il}\phi^{l}_{\ph{l}j}
%\bigg] \label{quintnewton}
%\end{align}
%

 Thus \emph{to Newtonian order there is no Vainshtein mechanism}; the canonical kinetic term will entirely determine the behaviour of $\phi$ to order $N=2$. 
We note that this result has already been noted in  \cite{Bloomfield:2014zfa} which shows consistency with out method.
In going beyond  \cite{Bloomfield:2014zfa}, we discuss below what happens when we consider post-newtonian corrections to higher orders in $N$.

\subsection{Effects to Post-Newtonian order}
It is natural to then wonder whether going to post-Newtonian order $N=4$ reveals non-linear behaviour for $\varphi^{PN} \equiv\varphi^{(4,V)}$ that was simply not there at Newtonian order. This may happen if to $N=4$, the scalar field equation possesses solutions describing an `inside' region where the dominant term is nonlinear in $\varphi^{PN}$. 
Indeed, to order $N=4$, the Quartic Galileon term present in the scalar equation does contribute. In that case, there will be new Vainshteinian potentials appearing
 to order $N=4$ and these will be different on each side of the Vainshtein radius. However, the effects of these potentials, being of higher order, will be to introduce
PN corrections to Brans-Dicke theory, rather than to GR, and the Vainshtein mechanism is inadequate for restoring GR around massive sources.
%However, by inspection of the full field equations, the scalar field to order $N=4$ is schematically of the form:

%\begin{align}
%\grad^{2}\varphi^{PN} &= {\cal S}(\varphi^{(2)},h^{(2)})
%\end{align}
%
%where ${\cal S}$ is some function of $\varphi^{(2)}$ and $h^{(2)}$, which are Newtonian order scalar field and metric fields respectively. Thus - as at Newtonian order - we find that the scalar field equation to $N=4$ post-Newtonian order is \emph{linear} in $\varphi^{PN}$ and so there is no non-linear regime for this field, and therefore no Vainshtein mechanism.

\section{Discussion and Conclusions}
\label{discussion}
In this paper we extended the work done in \cite{AvilezEtAl2015} by  applying the Parameterized Post-Newtonian-Vainshteinian formalism to the Quartic and Quintic Galileon theories. 
The PPNV formalism is an extension to the PPN formalism adapted to theories with an extra scale, within which non-linearities become important. In the case of the Galileon theories in question this scale is the  Vainshtein radius. The Vainshtein radius acts as the boundary between the inside and outside regions, each of which needs to be expanded independently in PPNV orders.  The inside region, is generally characterized by Vainshtein screening where non-linear kinetic terms predominantly determine the behaviour of the scalar field. The outside region, on the other hand, has dynamics for the scalar field dominated by the linear kinetic terms and the behaviour of the theory is approximately that of Brans-Dicke theory.

The PPNV formalism was constructed as a tool to facilitate constraining modified gravity theories (with extra scales) with Solar System and other strong field data.  In particular such constraints would be especially significant as they would be independent from cosmological constraints. Each theory to which one applies the PPNV formalism will produce a different set of potentials with corresponding coefficients. Identifying the form of these potentials and their coefficients, for a given theory, will therefore be the key to constrain it with available data and possibly may help direct the design of new experiments.  

In this work we focused on applying the PPNV formalism to the Quartic and Quintic Galileon. In both cases we expanded  both the inside and outside regions in PPNV orders. For the Quartic Galileon we also explicitly found the solution, up to PPNV order $(2, -3)$ outside and $(2,2)$ inside, in spherical symmetry. 
We confirmed previous works\cite{Burrage:2010rs,Bloomfield:2014zfa} that in the inside region there is a correction 
to the Newtonian potential proportional to $r$ while we found that the next correction comes to 
order $r^3$. This correction would produce an approximately constant force outside a massive body, that may be observed. The extra force produced by this extra correction is very small, however, by looking at systems like binary pulsars with observations integrated over a long period of time, this effect may still be within the reach of observation. Furthermore, it has been argued that certain Galileon models might produce observable effects in other `strong-gravity' systems \cite{Burrage:2010rs, Chagoya:2014fza}. %\CSC{It can be seen from (\ref{gamma4in}) that the strength of the modification to gravity within the Vainshtein radius is proportional to $r_{V}^{-2}$. }

The Quintic Galileon case, however displayed an interesting feature. To Newtonian order we found that there is no Vainshtein mechanism and therefore the fifth force produced by the kinetic term of $\phi$ will not be screened. This implies that to this order the constraints on the Quintic Galileon theory (\ref{quintjor}) on scales such as those of the solar system will be on the Brans-Dicke limit of the theory. 

One issue that may arise, is whether the inside and outside solutions can be matched, producing an approximate solution accross the Vainshtein radius. Unfortunately, this is not possible. If one tries to impose the matching conditions for the metric and its extrinsic curvature at $r=r_V$, one finds that no consistent matching
can be found. This is to be expected as the Vainshtein radius is a place of inherently non-perturbative behaviour. Fortunately, this is does not impose an obstacle.
One can use the solutions on either side to impose bounds on the parameters of the theory and this may be achievable by considering specific physical systems
which reside on each side.
 
The formalism developed here can be used in the case of other theories which exhibit kinetic screening, for instance, theories
 related to Modified Newtonian Dynamics\cite{Milgrom1983,BekensteinMilgrom1984},
 and their relativistic counterparts \cite{Bekenstein2004,Skordis2008,Skordis2009,BabichevDeffayetEsposito-Farese2011}
by appropriately extending the formalism in~\cite{Sagi:2010ei}.  It can also serve as a guide on how to design experimental
efforts for probing gravity in the solar system or in the regime of binary pulsars, see \cite{Sakstein:2017pqi}. All such ``strong-field'' systems will tend to push the Vainshtein radius $r_V$ to larger values.  On the cosmological side, cosmological constraints may place upper bounds on scales such as $r_{V}$ \cite{Renk:2017rzu}, thus, pushing $r_V$ to smaller values. If the two types of constraints become incompatible, then the theory is ruled out completely, unless $r_V$ is so small, that
even the solar system lies in the outside Vainshtein region, in which case constraints on Brans-Dicke theories apply.
 Indeed, if the Quartic Galileon is not to play a role of Dark Energy, although this has less immediate motivation,
 pushing $r_V$ to smaller values could evade gravitational
wave constraints \cite{Baker2017,Ezquiaga2017,SaksteinJain2017,CopelandEtAl2018} and in such case, ``strong-field'' systems will have an important role to play.
 Therefore, by combining both types of constraints there is the potential to create a zone of observational exclusion for such models.

\section*{Acknowledgements}
The research leading to these results 
has received funding from the European Research Council under the European Union's Seventh Framework 
Programme (FP7/2007-2013) / ERC Grant Agreement n.\,617656 ``Theories and Models of the Dark Sector: Dark Matter, Dark Energy and Gravity''.

% may be useful for beyond horndeski 
%\bibitem{NS2017} B.~P.~Abbott {\it et al.} [LIGO Scientific and Virgo and Fermi-GBM and INTEGRAL Collaborations], \href{http://iopscience.iop.org/article/10.3847/2041-8213/aa920c/meta}{Astrophys.\ J.\  {\bf 848} (2017) no.2,  L13} [\href{https://arxiv.org/abs/1710.05834}{astro-ph/1710.05834}]
%\bibitem{NS20171} B.~P.~Abbott {\it et al.} [LIGO Scientific and Virgo Collaborations], \href{https://journals.aps.org/prl/abstract/10.1103/PhysRevLett.119.161101}{Phys.\ Rev.\ Lett.\  {\bf 119} (2017) no.16,  161101} [\href{https://arxiv.org/abs/1710.05832}{gr-qc/1710.05832}].

\begin{widetext}
\appendix

\section{Field equations for the Quartic Galileon}
\label{eq_quartic}

We introduce the short hand notation $\phi_\mu \equiv \nabla_\mu \phi$, $\phi_{\mu\nu} = \nabla_\mu \nabla_\nu \phi$
and similarly for $Y$. We display the ordinary field equations in the usual sense as well as the dual field equations for which the dual field
$B$ such that $\nabla_\mu B = \alpha_4^{1/3} \nabla_\mu \phi$ is introduced (and $B_\mu = \nabla_\mu B$ is to be understood).

\subsection{Ordinary field equations}
The Einstein equations take the form
\begin{align}
  \phi \left( 1 +  \sigma_Q  Y^2 \right)  R^\mu_{\;\;\nu} 
&=  8\pi G \left(T^\mu_{\;\;\nu} -  \frac{1}{2} T \delta^\mu_{\;\;\nu} \right)
   + \frac{\omega}{\phi}  \phi^\mu   \phi_\nu   
+  \frac{1}{2}  \square \phi\delta^\mu_{\;\;\nu}
%\nonumber 
%\\
%& 
+   \phi^\mu_{\;\;\nu} 
+  \sigma_Q \bigg[ \At_1  \phi^\mu   \phi_\nu   +  \At_2     \delta^\mu_{\;\;\nu} 
+ \At_3 W^\mu_{\;\;\nu} 
\nonumber 
\\
& 
+  \At_4  \phi^\mu_{\;\;\nu} 
+ 2   \phi\left(    Y^\mu Y_\nu  - L^\mu_{\;\;\nu}\right)
%\nonumber 
%\\
%& 
 - 2  \phi Y   \left( F^\mu_{\;\;\nu}   +   N^\mu_{\;\;\nu}   +   Q^\mu_{\;\;\nu} \right)
\bigg]
%\nonumber 
%\\
%& 
%+  \sigma_Q \bigg[  Y \At_1    - 2\At_2     
%-  \At_3  I_\phi
%- \frac{1}{2} \At_4  \square \phi
%- 3   \phi I_Y  
% +   \phi Y   \left( F   +   3 Q \right)
%\bigg] \delta^\mu_{\;\;\nu}
\label{Quartic_galileon_Einstein_equations}
\end{align}

%\begin{align}
%\\
%  \phi \left( 1 +  \sigma_Q  Y^2 \right)  G^\mu_{\;\;\nu} 
%&= 8\pi G T^\mu_{\;\;\nu} 
%   + \frac{\omega}{\phi}  \phi^\mu   \phi_\nu   
%+  \left(   \frac{\omega}{\phi} Y  - \square \phi\right) \delta^\mu_{\;\;\nu}
%\nonumber 
%\\
%& 
%+   \phi^\mu_{\;\;\nu} 
%+  \sigma_Q \bigg[ \At_1  \phi^\mu   \phi_\nu   +  \At_2     \delta^\mu_{\;\;\nu} 
%+ \At_3 W^\mu_{\;\;\nu} 
%\nonumber 
%\\
%& 
%+  \At_4  \phi^\mu_{\;\;\nu} 
%+ 2   \phi\left(    Y^\mu Y_\nu  - L^\mu_{\;\;\nu}\right)
%\nonumber 
%\\
%& 
% - 2  \phi Y   \left( F^\mu_{\;\;\nu}   +   N^\mu_{\;\;\nu}   +   Q^\mu_{\;\;\nu} \right)
%\bigg]
%\label{Quartic_galileon_Einstein_equations}
%\end{align}
where
\begin{equation}
\sigma_Q =  \frac{M_p^2}{16\Lambda^6\phi^6}=   \frac{\alpha_4}{16\phi^6}
\end{equation}
and where  for the ease of avoiding long expresions we have defined the following:
\begin{align*}
I_\phi &= Y_\alpha \phi^\alpha 
& \qquad \qquad &
I_Y = Y_\alpha  Y^\alpha
\\
Z_\phi   &= I_Y  + \square\phi I_\phi 
 & \qquad \qquad &
V_\phi   =  Y \square\phi - I_\phi
\\
L_\mu  &= Y_\rho \phi^\rho_{\;\;\mu} 
 & \qquad \qquad &
L^\mu_{\;\;\nu}  = L^\mu \phi_\nu + \phi^\mu L_\nu
\\
Q_{\mu\nu} &=  R_{\mu\alpha\nu\beta} \phi^\alpha \phi^\beta
 & \qquad \qquad &
Q_\alpha = R_{\alpha\beta}   \phi^\beta
\\
Q &= Q_{\alpha} \phi^\alpha
 & \qquad \qquad &
W^\mu_{\;\;\nu}  = Y^\mu \phi_\nu + \phi^\mu Y_\nu
\\
N^\mu_{\;\;\nu}  &= Q^\mu \phi_\nu + \phi^\mu Q_\nu
 & \qquad \qquad &
F_{\mu\nu} =  \phi^\rho_{\;\;\mu}  \phi_{\rho\nu}
\\
F &= F^\mu_{\;\;\mu}
 & \qquad \qquad &
J_\phi =  (\square \phi)^2 - F
\\
U_\phi &= 
(\square \phi)^3 - 3\phi^\alpha_{\;\;\beta} \phi^{\beta}_{\;\;\alpha} \square \phi +2\phi^\alpha_{\;\;\beta} \phi^\beta_{\;\;\gamma}\phi^\gamma_{\;\;\alpha}
 =  \square\phi[ (\square \phi)^2  - 3 F ] + 2 F^\mu_{\;\;\nu} \phi^\nu_{\;\;\mu}
\\
\At_1 &=   R  \phi Y    +   \phi   J_\phi -15  Y  \square\phi +  \frac{63}{2} \frac{1}{\phi}  Y^2
%\\
%%\At_2 &=    \frac{81}{2} \frac{1}{\phi}  Y^3  + 2 \phi Y    Q - \phi Y     J_\phi -2   \phi  Z_\phi  - 15  Y V_\phi  
\\
 \At_2  &=    R  \phi Y^2   
 +   \phi  Y (J_\phi  +Q)
- \frac{9}{\phi}  Y^3
- \frac{15}{2} Y^2  \square \phi  
-    \phi  \square \phi   I_\phi 
-   \phi I_Y  
\\
 \At_3 &=    - 15 Y  +  2 \phi  \square \phi  
\\
 \At_4 &=  2 \phi Y   \square \phi +  2  \phi  I_\phi + 15  Y^2  
\end{align*}
The scalar field equation for this theory takes the form
\begin{align}
 &  
  2\omega \left( \square\phi + \frac{Y}{\phi} \right) +   \phi R
+\sigma_Q \bigg\{ 
  315 \frac{Y^3}{\phi}  
+  126 Y I_\phi 
-  117   Y^2 \square \phi  
+ 30  \phi\left( Y Q -  Z_\phi \right)
\nonumber
\\
 & 
+ 15 Y^2 \phi R  
 + 2 \phi^2  \bigg[ U_\phi 
-4 Y^\mu Q_\mu
%\nonumber
%\\
% & 
 +  R  I_\phi  -2 \square \phi Q +2  Q_{\mu\nu} \phi^{\mu\nu} 
-2  G^{\mu\nu} \phi_{\mu\nu}   Y 
\bigg] 
\bigg\}
= 0
\label{Quartic_galileon_scalar_equation}
\end{align}

\subsection{Dual field equations}
\label{app_dual}
Defining $\beta = B_\mu B^\mu$ and
\begin{align*}
 \tilde{I}_\phi &= \tilde{Y}_\mu B^\mu = \alpha^2 I_\phi
 & \qquad \qquad &
 \tilde{I}_Y = \tilde{Y}_\mu  \tilde{Y}^\mu =  \alpha^{8/3} I_Y
\\
\tilde{Z}_\phi &= \tilde{I}_Y + \square B \tilde{I}_\phi = \alpha^{8/3}Z_\phi
& \qquad \qquad &
\tilde{V}_\phi =  -\frac{1}{2} \beta \square B - \tilde{I}_\phi =  \alpha^2  V_\phi
\\
\tilde{L}_\mu  &= \tilde{Y}_\rho B^\rho_{\;\;\mu}  = \alpha^2 L_\mu
& \qquad \qquad &
\tilde{L}_{\mu\nu} = 2 \tilde{L}_{(\mu} B_{\nu)}  = \alpha^{8/3} L_{\mu\nu}
\\
  \tilde{Q}_{\mu\nu} &= R_{\mu\alpha\nu\beta}  B^\alpha B^\beta = \alpha^{4/3} Q_{\mu\nu} 
& \qquad \qquad &
 \tilde{Q}_\mu = R_{\mu\nu} B^\nu=  \alpha^{2/3} Q_\mu
\\
 \tilde{Q} &= \tilde{Q}_\mu  B^\nu=  \alpha^{4/3}  Q
& \qquad \qquad &
\tilde{W}_{\mu\nu} = 2\tilde{Y}_{(\mu} B_{\nu)} =  \alpha^2  W_{\mu\nu}
\\
\tilde{N}_{\mu\nu} &= 2 B_{(\mu} \tilde{Q}_{\nu)} = \alpha^{4/3} N_{\mu\nu}
& \qquad \qquad &
 \tilde{F}_{\mu\nu} = B_{\rho(\mu} B^\rho_{\;\;\nu)} = \alpha^{4/3} F_{\mu\nu}
\\
 \tilde{F} &= \tilde{F}^\mu_{\;\;\mu} = \alpha^{4/3} F
& \qquad \qquad &
 \tilde{J}_\phi = (\square B)^2 - \tilde{F} = \alpha^{4/3} J_\phi
\\
 \tilde{Y}_\mu &= - B^\nu B_{\mu\nu} = \alpha^{4/3} Y_\mu
\\
 \tilde{U}_\phi &= (\square B)^3 - 3 \square B  B^{\mu\nu} B_{\mu\nu} + 2  B^{\mu\nu}  B_{\nu\rho}  B^\rho_{\;\;\mu}  = \alpha^2 U_\phi
\end{align*}
the dualized Einstein equationas are
\begin{align}
  \phi \left( 1 +   \frac{1}{64\alpha^{2/3}} \frac{\beta^2}{\phi^6} \right)  R^\mu_{\;\;\nu} 
&= 8\pi G\left( T^\mu_{\;\;\nu}  - \frac{1}{2} T \delta^\mu_{\;\;\nu}\right)
 +   \frac{\omega}{\alpha^{4/3}} \frac{1}{\phi}  B^\mu   B_\nu   
 +  \frac{1}{\alpha^{2/3}} \left(  B^\mu_{\;\;\nu} +    \frac{1}{2}   \square B   \delta^\mu_{\;\;\nu} \right)
 +  \frac{1}{16\phi^6} \bigg\{   \hat{A}_1  B^\mu   B_\nu 
\nonumber 
\\
&  
 +  \hat{A}_2     \delta^\mu_{\;\;\nu} 
+ \hat{A}_3 \tilde{W}^\mu_{\;\;\nu} 
+   \hat{A}_4  B^\mu_{\;\;\nu} 
+  \frac{1}{\alpha^{2/3}}  \phi \left[ 
2 \tilde{Y}^\mu \tilde{Y}_\nu  -2 \tilde{L}^\mu_{\;\;\nu} 
+   \beta  \left( \tilde{F}^\mu_{\;\;\nu}   +  \tilde{N}^\mu_{\;\;\nu}   +   \tilde{Q}^\mu_{\;\;\nu} \right) \right]
\bigg\}
\label{Quartic_galileon_dual_Einstein_equations}
\end{align}
where 
\begin{align}
\hat{A}_1 &=  \frac{1}{\alpha^{2/3} } \left[
 \phi   \tilde{J}_\phi  - \frac{1}{2}  \beta  \phi R  +   \frac{15}{2\alpha^{2/3}}  \beta  \square B 
 +    \frac{63}{8\alpha^{4/3}}  \frac{\beta^2}{\phi} \right]
=  \alpha^{2/3}  \At_1
\\
\hat{A}_2 &=   
\frac{1}{\alpha^{2/3}} \left[
 \frac{9}{8\alpha^{4/3}}  \frac{\beta^3}{\phi} 
-  \phi \tilde{I}_Y 
- \frac{15}{8\alpha^{2/3}}  \beta^2 \square B 
-  \phi  \tilde{I}_\phi\square B
- \frac{1}{2}   \phi  \beta (\tilde{J}_\phi  + \tilde{Q}  )
 + \frac{1}{4}  \beta^2  \phi R 
\right] = \alpha^2 \At_2 
\\
 \hat{A}_3 &=     \frac{1}{\alpha^{2/3}} \left[   2 \phi   \square B  + \frac{15}{2\alpha^{2/3}}  \beta \right] = \At_3
\\
 \hat{A}_4 &=  \frac{1}{\alpha^{2/3}} \left[  -\beta \phi  \square B
   +  2 \phi  \tilde{I}_\phi + \frac{15}{4\alpha^{2/3}}  \beta^2   \right] = \alpha^{4/3}   \At_4
\end{align}
and the dual scalar field equation is
\begin{align}
 &  
   \phi R + \frac{1}{8\phi^4}   \left[ \tilde{U}_\phi -4 \tilde{Y}^\mu \tilde{Q}_\mu +  R \tilde{I}_\phi  - 2 \tilde{Q}  \square B 
  + 2 \tilde{Q}_{\mu\nu} B^{\mu\nu} +  \beta  G^{\mu\nu} B_{\mu\nu}   \right] 
+ \frac{ 2\omega}{\alpha^{2/3}} \left(   \square B    -\frac{1}{2\alpha^{2/3}}  \frac{\beta }{\phi} \right) 
\nonumber
\\
 & 
+   \frac{1}{16\alpha^{2/3}} \frac{1}{\phi^6} \bigg\{ 
 \frac{15}{4}  \beta^2 \phi R  
- 15 \phi\left( \beta \tilde{Q} +  2\tilde{Z}_\phi \right)
-  \frac{63}{\alpha^{2/3}}  \beta \tilde{I}_\phi 
- \frac{117}{4\alpha^{2/3}}  \beta^2  \square B   
 - \frac{315}{8\alpha^{4/3}}  \frac{\beta^3}{\phi}
\bigg\}
= 0
.
\label{Quartic_galileon_dual_scalar_equation}
\end{align}

\section{Field equations for the Quintic Galileon}
\label{eq_quintic}
The Einstein equations are
\begin{align}
&
\phi \left(1- 36  \sigmaQui   \frac{Y^3}{\phi^2}   \right) R^\mu_{\;\;\nu}
 = 8\pi G\left( T^\mu_{\;\;\nu} - \frac{1}{2}  T  \delta^\mu_{\;\;\nu}  \right)
+  \frac{\omega}{\phi} \phi^\mu   \phi_\nu  
+ \phi^\mu_{\;\;\nu} + \frac{1}{2}   \square \phi \delta^\mu_{\;\;\nu} 
\nn
\\
&
+ \sigmaQui  \bigg\{
 \tilde{B}_1 \delta^\mu_{\;\;\nu} 
+ \tilde{B}_2 \phi^\mu \phi_\nu  
+ \tilde{B}_3  \phi^\mu_{\;\;\nu} 
+ \tilde{B}_4  W^\mu_{\;\;\nu} 
+ 6 Y \left(   4 \frac{Y}{\phi}   -   \square \phi \right) \left(N^\mu_{\;\;\nu} + Q^\mu_{\;\;\nu}   \right)
 + 6 \left( \square\phi  -    \frac{Y}{\phi}  \right) \left(Y^\mu Y_\nu  -  L^\mu_{\;\;\nu} \right)
\nn
\\
&
+    6   \left(   C^\mu \phi_\nu  +  \phi^\mu C_\nu  \right)
  - 6   (Y^\mu L_\nu + L^\mu Y_\nu) 
- 6 \left(    10   \frac{Y^2}{\phi} +2 Y \square \phi        +  I_\phi   \right) F^\mu_{\;\;\nu}  
+ 6   Y  \bigg[
    F^\mu_{\;\;\alpha}   \phi^\alpha_{\;\;\nu} +\phi^\mu_{\;\;\alpha} F^\alpha_{\;\;\nu}   
 - S^\mu_{\;\;\nu}
-  R^\mu_{\;\;\alpha\nu\beta} W^{\alpha\beta}
\nn
\\
& 
- Y^\rho R^\mu_{\;\;\rho} \phi_\nu
- Y^\rho \phi^\mu R_{\rho\nu} 
+ I_\phi R^\mu_{\;\;\nu} 
-  Q^\mu Y_\nu
-  Y^\mu Q_\nu 
+ Q^\mu_{\;\;\rho} \phi^\rho_{\;\;\nu}
+\phi^\mu_{\;\;\rho} Q^\rho_{\;\;\nu} 
 + Q_\rho \phi^{\rho\mu}  \phi_\nu
+ Q_\rho \phi^\rho_{\;\;\nu}  \phi^\mu
\bigg]
\bigg\}
\label{eq_quintic_ordinary_Einstein}
\end{align}
where
\begin{equation}
 \sigmaQui =   \frac{\alpha_5}{32} \frac{1}{\phi^7} 
\end{equation}
and where  for the ease of avoiding long expresions we have defined the following:
\begin{align}
S_\mu &= \phi^{\alpha\beta}\phi_\rho R^\rho_{\;\;\alpha\beta\mu} 
 & \qquad \qquad &
S^\mu_{\;\;\nu} = S^\mu \phi_\nu + \phi^\mu S_\nu
\\
C_\mu &=  L_\beta    \phi^\beta_{\;\;\mu}
 & \qquad \qquad &
I_L = Y_\mu  L^\mu
\end{align}
and
\begin{align}
\tilde{B}_1 &=
- 12 \frac{Y^3}{\phi} R 
- 3 Y I_\phi R
- 6  Y^2 G^\alpha_{\;\;\beta} \phi^\beta_{\;\;\alpha} 
  -    24  \frac{Y^2}{\phi}  J_\phi 
+ 24  \frac{Y}{\phi}  \square\phi   I_\phi  
  + 36 \frac{Y^4  }{\phi^3} 
+ 21  \frac{Y^3}{\phi^2}  \square\phi 
- 39  \frac{Y}{\phi}  I_Y
 -12 \frac{Y^2}{\phi}  Q 
\nn
\\
&
  -6  I_Y \square\phi  
+  6 Y \left( 2  Y^\alpha  Q_\alpha  + \square\phi Q \right) 
+ 6 Y S_\mu \phi^\mu 
 -3 I_\phi  J_\phi 
+  6 I_L
\\
\tilde{B}_2 &= 69  \frac{Y^2}{\phi^2}  \square\phi - 78   \frac{Y^3  }{\phi^3} - 24   \frac{Y}{\phi}  J_\phi     
- 12 \frac{Y^2}{\phi} R - 6  Y G^\alpha_{\;\;\beta} \phi^\beta_{\;\;\alpha} +    U_\phi  
\\
\tilde{B}_3 &= 6 \left[ 10  \frac{Y^2}{\phi}  \square\phi - 8  \frac{Y}{\phi}   I_\phi  - 16  \frac{Y^3}{\phi^2}   +  Y J_\phi -   Y Q +  Z_\phi \right] 
\\
\tilde{B}_4 &= 3 \left[ 23   \frac{Y^2}{\phi^2}  +   Y R    - 16  \frac{Y}{\phi}   \square\phi +  J_\phi \right]
\end{align}
The scalar field equation takes the form
\begin{align}
 (2\omega  +3) \square\phi 
 &= 
  8\pi G T
\nn
\\
&
+  \sigmaQui \bigg\{
  540   \frac{Y^4  }{\phi^3}
- 450   \frac{Y^3  }{\phi^2} \square\phi 
+ 330  \frac{Y^2  }{\phi^2} I_\phi 
+ 114  \frac{Y^2  }{\phi}   Q
- 102   \frac{Y}{\phi} I_Y
 + 138    \frac{Y^2}{\phi}  J_\phi 
\nonumber
\\
& 
- 228   \frac{Y}{\phi} \square\phi I_\phi  
+ 96  I_Y \square \phi 
+ 48 I_\phi J_\phi 
 - 96 I_L
+ 6 Y \left( \phi  J_\phi  + 8  I_\phi  + 14   \frac{Y^2}{\phi}  \right) R 
\nn
\\
&
- 12 Y  U_\phi 
+ 6\phi  J_\phi Q
- 96 Y \left( 2    Q^\mu Y_\mu - Q_{\mu\nu} \phi^{\mu\nu} +   \square \phi Q \right) 
- 12 \phi Q^\mu \left(  2   L_\mu - 2 \square\phi  Y_\mu + Y  Q_\mu \right)
\nonumber 
\\ 
& 
- 12G^{\mu\nu}\left[ 
 \phi  \left( 2 Y F_{\mu\nu} + Y Q_{\mu\nu} - 2 Y \square\phi  \phi_{\mu\nu} - Y_\mu Y_\nu -  I_\phi  \phi_{\mu\nu} \right)
 - 11  Y^2\phi_{\mu\nu}
\right]
\nn
\\
&
- 6 \phi \left[ 4 R_{\alpha\mu\beta\nu} \phi^{\mu\nu} \phi^\alpha Y^\beta +2YR_{\alpha\mu\beta\nu} \phi^{\mu\nu} \phi^{\alpha\beta}
- Y R_{\mu\rho\alpha\beta} R^{\nu\rho\alpha\beta} \phi^\mu \phi_\nu 
+2Q^{\mu\nu} \left( \square\phi  \phi_{\mu\nu} - F_{\mu\nu}   \right)
\right]
\nn
\\
&
- 2\phi  \left[ (\square \phi)^4  - 6  (\square \phi)^2  F + 3 F^2 +  8\square\phi F^{\mu\nu}\phi_{\mu\nu}   - 6 F^{\mu\nu} F_{\mu\nu} \right] 
\bigg\}
\label{eq_quintic_ordinary_scalar}
\end{align}

\end{widetext}

\bibliographystyle{unsrtnat}
\bibliography{references}

\end{document}